# Mapping and Spectroscopy of the Planetary Nebula NGC 7009 in the Visual and Infrared

J.P. Phillips[1], L.C. Cuesta[2], G. Ramos-Larios[1]

[1]Instituto de Astronomía y Meteorología, Av. Vallarta No. 2602, Col. Arcos Vallarta, C.P. 44130 Guadalajara, Jalisco, México   e-mail : jpp@astro.iam.udg.mx

[2]Centro de Astrobiología,  Instituto Nacional de Técnica Aeroespacial, Carretera de Ajalvir, km 4, 28850 Torrejón de Ardoz, Madrid, Spain
e-mail : cuestacl@inta.es

**Abstract**

NGC 7009 is a fascinating example of a high excitation, elliptical planetary nebula (PN) containing circum-nebular rings, and FLIERs and jets along the major axis. We present visual spectroscopy along multiple position angles through the nucleus, taken with the Observatorio Astronomico Nacional (Mexico); mid-infrared (MIR) spectroscopy and imaging acquired using the Infrared Space Observatory (ISO) and Spitzer Space Telescope (SST), and narrow band imaging obtained using the Hubble Space Telescope (HST). The data show that the mid-infrared (MIR) continuum is dominated by a broad $\approx$100K continuum, and a strong excess attributable to crystalline silicate emission. The primary peaks in this excess are similar to those observed in Forsterite and clino- and ortho-enstatite. The MIR images, by contrast, appear to be dominated by ionic transitions, with the $\lambda 8.991$ $\mu$m transition of [ArIII] being important in the 8.0 $\mu$m band. The morphology and size of the envelope are found to vary with wavelength, with the largest dimensions occurring at 8.0 $\mu$m – a trend which is also reflected in an increase in the 8.0$\mu$m/4.5$\mu$m and 5.8$\mu$m/4.5$\mu$m ratios with distance from the nucleus. The visual spectroscopy permits us to map density and temperature throughout the shell, and confirm that the lowest values of $n_e$ are located close to the ansae, where densities appear to be of order 900-2600 cm$^{-3}$. We provide mean line intensities for 116 transitions in six regions of the shell, and use mapping to confirm a systematic increase in excitation in the outer portions of the envelope. We finally use the ground-based spectroscopy, and ratioing of HST images to investigate the presence of shocks in the ansae and interior envelope. It is concluded that line ratios in the ansae may be partially consistent with shock excitation, although these features are primarily dominated by photo-ionisation. We also note evidence for shock excitation at the limits of the interior elliptical shell, and for multiple bow-shock structures centered upon the ansae. The orientations of the easterly bow-shocks may have varied over time, indicating precession of the collimating engine at a rate of ~5 10$^{-2}$ deg/yr, whilst the outward splaying of the westerly "jet" appears consistent with shock refraction modeling. We finally note that HST observations of the halo rings show them to have widths of the order of ~1-3 arcsec, and steep changes in surface brightness consistent with local shock activity.

**Key Words:** (ISM:) planetary nebulae: individual: NGC 7009 --- ISM: jets and outflows --- infrared: ISM --- (ISM:) dust, extinction

## 1. Introduction

NGC 7009 is an elliptical, high excitation planetary nebula (PN) having several unusual characteristics. It is clear for instance that the O/C ratio is significantly greater than unity (e.g. Hyung & Aller 1995a; Liu et al. 1995; Rubin et al. 1997). However, heavy element abundances derived using recombination lines (principally OII) appear to be significantly larger than those derived using collisionally excited transitions (Liu et al. 1995; see also Barker 1983; Kaler 1986; Mathis 1995; Rubin et al. 1997). It has been argued that such disparities may be attributable to fluctuations in the electron temperature $T_e$, and evidence for such variations has been noted in the Keck HIRES spectroscopy of Hyung, Fletcher & Aller (2003). However, the latter results conflict with the HST observations of Rubin et al. (2002), which suggest a variation in temperatures $t_A^2 \cong 3.5\ 10^{-3}$, and with the Argus array results of Tsamis et al. (2008) (which imply $t_A^2 \cong 7.5\ 10^{-4}$). Neither of these values is large enough to explain the discrepancy in abundances noted above.

One problem with these estimates is that they represent averages along the line of sight, however, and may underestimate the intrinsic range of the temperature fluctuations (Rubin et al. 2002). Alternatively, it has been suggested that the OII recombination lines may probe H-poor (oxygen rich) plasmas, which take the form of denser clumps or filaments (Liu et al. 2000; Tsamis et al. 2004). Barlow et al. (2006) point out that the narrowness of the OII lines (compared to [OIII]) tends to favour this hypothesis.

The source also represents one of the small fraction of nebulae showing evidence for ansae at the limits of the major axis, interpreted by Balick et al. (1998) as representing bow-shock structures. Such a presumption is also supported by their radial velocities of ~6.1-6.2 km s$^{-1}$ (Reay & Atherton 1985; Fernandez, Monteiro & Schwarz 2004), and mean proper motions of order ~28 mas yr$^{-1}$ (Fernandez, Monteiro & Schwarz 2004; Rodríguez & Gómez 2007), which suggest outflow velocities ~114 km s$^{-1}$ for a distance of 0.86 kpc (Fernandez, Monteiro & Schwarz 2004). This value is highly supersonic, and somewhat larger than the estimate of Reay & Atherton (1985) (~80(D/kpc) km s$^{-1}$).

It has additionally been suggested that levels of [NII] emission within the ansae are extraordinarily high, perhaps indicating abundances of nitrogen which are anomalously large (Balick et al. 1994; Czyzak & Aller 1979); a possible consequence of the dredge-up of materials in the central star progenitor (see e.g. Renzini & Voli 1981; Marigo et al. 1998; van den Hoek & Groenewegen 1997; Marigo 2008). However, recent spectroscopy and modelling have suggested that N abundances are normal, and that the high [NII] line intensities may be attributable to ionisation stratification (Gonçalves et al. 2003, 2005). It also appears that neither the jets nor ansae have ionic line ratios (principally [SII]/H$\alpha$ and [OIII]/H$\alpha$) which would be consistent with shock excitation of the local nebular shell (Gonçalves et al. 2003) – a problem which would appear, at first sight, to be in conflict with the bow-shock mechanism cited above.

Such ansae are often grouped together with the Fast Low Ionisation Emission Regions (FLIERs) (see e.g. Balick et al. 1993), and it is worth noting that further examples of such FLIERs have been noted in the inner regions of the source, towards the major axis limits of the bright elliptical shell (see e.g. Balick et al. 1994). This would be consistent with deduced outflow velocities for these features of ~60 km s$^{-1}$ (Reay & Atherton 1985), implying velocities which are highly supersonic. However, these structures appear to be oriented along a differing axis to that of the ansae, and it has been argued that they may represent low excitation shadows associated with compact condensations (Sabbadin et al. 2004).

It has finally been noted that a good fraction of PNe having FLIER-like structures also appear to be possession of annular rings within the extended (progenitor mass-loss) envelopes (e.g. Phillips et al. 2009). NGC 7009 is a typical representative of this class, with up to six rings having being discovered with a spacing of between 2.9 and 4.8 arcsec (Corradi et al. 2004; see also Balick et al. (1998) for a "pre-discovery" image of these rings). However, the characteristics of such rings, and the mechanisms responsible for their formation are still far from being well established (see e.g. Phillips et al. 2009; Corradi et al. 2004, and references therein).

Finally, we note that stellar wind mass-loss rates remain extremely uncertain, with estimates ranging from 3 10$^{-10}$ M$_\odot$ yr$^{-1}$ to as high as 10$^{-7}$ M$_\odot$ yr$^{-1}$ (Sonneborn et al. 2008; Cerruti-Sola & Perinotto 1985, 1989; Bombeck et al. 1986; Hutsemekers & Surdej 1989; Tinkler & Lamers 2002); that the distance is far from well established, with estimates ranging from 0.6 and 2.1 kpc (see e.g. the critical discussion of Sabbadin et al. 2004); and that estimates of central star temperature vary from 8.2 10$^4$ to 9.5 10$^4$ K (Méndez et al. 1992; Kingsburgh & Barlow 1992; Sabbadin et al. 2004;

Sonneborn, Iping & Herald 2009), an uncertainty which is of critical importance in modelling the spectrum of the source.

It is therefore clear that although NGC 7009 represents one of the most extensively studied of Galactic PNe, much of our information concerning this source appears to be contradictory or uncertain; a consequence of inaccurate or partial observational results, and lack of understanding of the nebular emission and formation processes.

We shall present new visual spectroscopy of the source taken with a Boller & Chivens spectrograph, in which mean line intensities are derived for seven position angles through the nucleus, and six differing regions of the nebular shell. We shall also present Spitzer and ISO spectroscopy of the nebular envelope, in which we find evidence for a warm ($T_{GR} \approx 10^2$ K) dust continuum and silicate emission features. Mapping at 3.6, 4.5, 5.8 and 8.0 $\mu$m will be found to indicate significant changes in source morphology and size, and variations in colours for differing sectors of the nebular shell, whilst HST imaging is used to illustrate the variations of structure for differing ionic ratios; define the morphology of the ansae bow shock structures; and explore the role of shock excitation in creating and defining these features. We shall also note the presence of a likely shock interface at the limits of the interior elliptical shell, and note that the halo rings appear non-circular, and of variable width. It is suggested that steep changes in their surface brightnesses may imply local shock activity.

## 2. Observations

We shall be making use of data products (imaging and spectroscopy) deriving from SST programs 93 (Survey of PAH Emission, 10-19.5 $\mu$m) and 68 (Studying Stellar Ejecta on the Large Scale using SIRTF-IRAC); spectroscopy obtained using the long and short wave spectrometers of the Infrared Space observatory (ISO)[1]; Hubble Space telescope (HST) imaging with the Wide Field Planetary Camera 2 (WFPC2); and data deriving from long slit spectroscopy with a Boller and Chivens Spectrograph.

---

[1] *Based on observations with ISO, an ESA project with instruments funded by ESA Member States (especially the PI countries: France, Germany, the Netherlands and the United Kingdom) and with the participation of ISAS and NASA.*

Spectroscopy of NGC 7009 was obtained as a result of SST Program 93 (Survey of PAH Emission, 10-19.5 um) undertaken on 14/05/2004. This resulted in a 9.9-19.6 $\mu$m spectrum of the central portions of the source ($\alpha$ = 316.0447°; $\delta$ = -11.3637°), using the Short-High (SH) module of the Infrared Spectrograph (IRS; Houck et al. 2004). Similarly, we make use of spectroscopy deriving from the ISO, a 60 cm f/15 telescope which was launched on 17/11/1995. This carried a Short Wave Spectrometer (SWS; de Grauuw et al. 1999) operating between 2.4 and 45 $\mu$m, and a Long Wave Spectrometer (LWS; Clegg et al. 1999) operating from 45 to 198.8 $\mu$m. Further details of the apertures of these instruments are provided in Sect. 3. Both sets of spectra were taken on 23/11/1997, were centred on $\alpha$ = 24h 04 m 10.80 s, $\delta$ = -11° 21' 56.6", and had exposure periods of 1912 s (SWS) and 1268 s (LWS).

Spitzer observations of NGC 7009 with the Infrared Array Camera (IRAC; Fazio et al. 2004) took place on 27/10/2004 as part of the program 68, and have been processed as described in the IRAC instrument handbook (Carey et al. 2010). The resulting post- Basic Calibrated Data (post-BCD) are relatively free from artefacts; well calibrated in units of MJy sr$^{-1}$; and have reasonably flat emission backgrounds. The observations employed filters having isophotal wavelengths (and bandwidths $\Delta\lambda$) of 3.550 $\mu$m ($\Delta\lambda$ = 0.75 $\mu$m), 4.493 $\mu$m ($\Delta\lambda$ = 1.9015 $\mu$m), 5.731 $\mu$m ($\Delta\lambda$ = 1.425 $\mu$m) and 7.872 $\mu$m ($\Delta\lambda$ = 2.905 $\mu$m). The normal spatial resolution for this instrument varies between ~1.7 and ~2 arcsec (Fazio et al. 2004), and is reasonably similar in all of the bands, although there is a stronger diffraction halo at 8 $\mu$m than in the other IRAC bands. This leads to differences between the point source functions (PSFs) at ~0.1 peak flux.

We have used these data to produce colour coded combined images of the sources in three IRAC bands, where 3.6 $\mu$m is represented as blue, 4.5 $\mu$m as green, and 8.0 $\mu$m is indicated as red (Fig. 1). These results, as well as the HST images to be described below, have also been processed using unsharp masking techniques, whereby a blurred or "unsharp" positive of the original image is combined with the negative. This leads to a significant enhancement of higher spatial frequency components, and an apparent "sharpening" of the image (see e.g. Levi 1974). We have additionally produced contour mapping, and band ratio profiles and mapping, details of which are provided in Sect. 4.

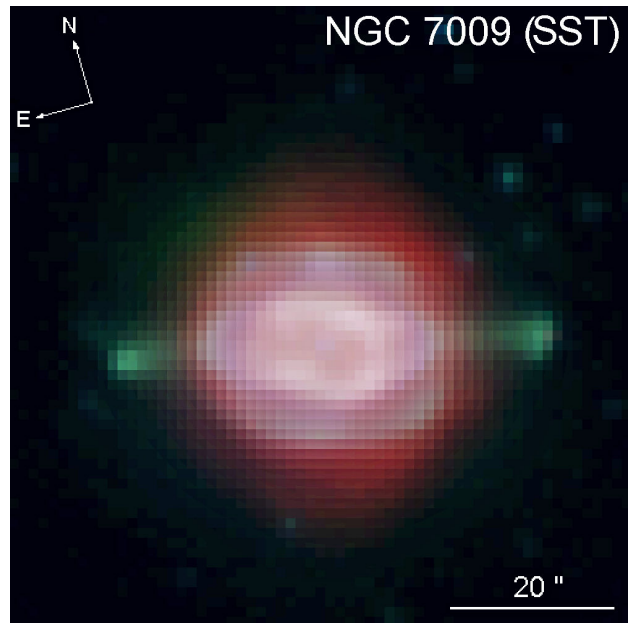

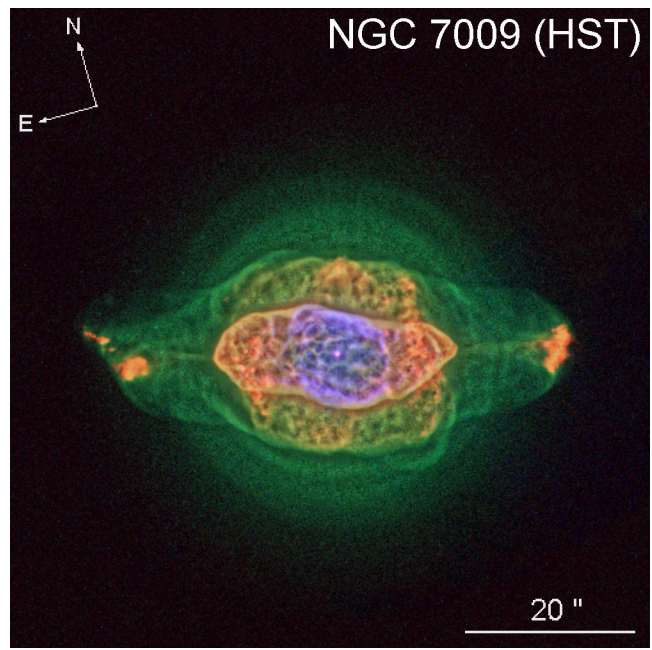

**Figure 1:** Images of NGC 7009 in the mid-infrared (upper panel) and visual (lower panel). In the former case, we have combined IRAC images taken with the SST in the 3.6 μm (blue), 4.5 μm (green) and 8.0 μm (red) bands. It will be noted that the ansae and jets have a predominantly green colouration, indicating higher levels of emission in the 4.5 μm band. It also appears that the central elliptical shell is surrounded by a redder, more circular halo, indicating higher levels of emission in the 8.0 μm band. The lower panel, by contrast, represents a composite HST image, in which HeII λ4686 corresponds to blue, [OIII] λ5007 to green, and [NII] λ6583 is represented by red. Both of these images have been processed using unsharp masking.

Optical spectroscopy of NGC 7009 was obtained on 19-21 September 2000, using the 2.1 m telescope at the Observatorio Astronómico Nacional (OAN) at San Pedro Mártir, Baja California, México. The Boller & Chivens Spectrograph had a grating of 600 $\ell$ mm$^{-1}$, a blaze of 4550 Å, and a total slit length of 160 arcsec. The spectra were recorded using a Thomson 2K (CCD2000) 2048x2048 pixel$^2$ TH7398M charge-coupled device (CCD), with pixel sizes of 15x15 $\mu$m$^2$, and calibrated using a He/Ar Lamp. This resulted in a usable wavelength range of $\Delta\lambda$ = 2500 Å (i.e. between 4330 and 6830 A), a plate scale of 1.54 Å pixel$^{-1}$, and a spectral resolution of 3.9 Å. The exposures used for the line strength estimates and mapping were in all cases taken through nucleus of the source, and along position angles 0°, 30°, 60°, 76°, 90°, 120°, 150°. The exposure periods along 0° and 90° were respectively 600 and 900 s, with the other PAs being exposed for 300 s. Two other longer exposures (of 900 s) were also taken north and south of the centre (i.e. along non-radial directions), and these were used to provide checks on relative line intensities. Finally, flux calibration was undertaken using the star HD217086 (Massey et al. 1988), and the results analysed using the FIGARO data reduction package (Shortridge 1993). Atmospheric extinction was corrected using a curve published at the OAN website. All of the OAN telescopes suffered from light leakage over this period, resulting in contamination from scattered and reflected moonlight. The present results were taken during a grey/dark period, however, and levels of contamination are modest – differences in sky/reflection background vary by less than 3% between one side of the slit and the other. These gradients were removed by assessing mean backgrounds at the spatial limits of the exposure, and removing them from the spectrum.

The HST was launched in April 1990, and consists of a f/24 2.4 m Ritchey-Chretien reflector[2]. The observations of NGC 7009 were obtained using the WFPC2 (Holtzman et al. 1995; see also the WFPC2 Instrument Handbook (Biretta 1996)), took place over a variety of dates ranging between 1996 and 2000, and correspond to HST programs 6117 (FLIERs in Planetary Nebulae; principal investigator (P.I.): B. Balick) and 8114 (Fundamental Problems in Astrophysics: The Thermal Equilibrium of NGC 7009; P.I.: R.H. Rubin). The enhanced imaging results were downloaded from the Hubble Legacy Archive at http://hla.stsci.edu/, and are astrometrically corrected, and in units of electrons/sec. We have selected exposures taken with filter F502N (pivot wavelength $\lambda_p$ = 5012.4 Å, effective bandwidth $\Delta\lambda$ = 26.9 Å) corresponding to emission from the $\lambda$5007 transition of [OIII], and a total exposure time of $\Delta t$ = 40 s; filter F658N ($\lambda_p$ = 6590.8 Å, $\Delta\lambda$ = 28.5 Å, $\Delta t$ = 1200 s, [NII] $\lambda$6584 Å); filter F469N ($\lambda_C$ = 4694.3 Å, $\Delta\lambda$ = 25.0 Å, $\Delta t$ = 160 s, HeII $\lambda$4686 Å); filter F656N ($\lambda_C$ = 6563.8 Å, $\Delta\lambda$ = 21.5 Å, $\Delta t$ = 100 s, H$\alpha$ $\lambda$6563 Å); and filter 673N ($\lambda_C$ = 6732.2 Å, $\Delta\lambda$ = 47.2 Å, $\Delta t$ = 900 s, [SII] $\lambda$6717+6732 Å). All of these latter exposures were acquired though HST proposal 6117. We have, in addition, made use of the deeper ($\Delta t$ = 320 s) [OIII] exposure acquired in April of 2000, as part of proposal 8114. The spatial resolution is 0.10 arcsec/pixel.

---

[2] *Based on observations made with the NASA/ESA Hubble Space Telescope, and obtained from the Hubble Legacy Archive, which is a collaboration between the Space Telescope Science Institute (STScI/NASA), the Space Telescope European Coordinating Facility (ST-ECF/ESA) and the Canadian Astronomy Data Centre (CADC/NRC/CSA).*

One of the images (the longer [OIII] exposure) shows evidence for the annular features described by Corradi et al. (2004), as well as possible internal shock rims. We have obtained profiles through the rings by taking a traverse across the nucleus, fitting the underlying (and smoother) component of emission using a fourth order least-squares polynomial fit, and removing this from the total surface-brightness fall-off. This permits us to investigate variations in emission due to the ring structures alone. We have also investigated nebular excitation and shocks by obtaining ratios between the images, and investigated the mechanisms responsible for ansa emission in colour-colour diagnostic planes. This process is described in fuller detail in Sects. 6 & 7.

## 3. ISO and Spitzer Spectroscopy of NGC 7009

An ISO spectrum of NGC 7009 is illustrated in Fig. 2, where there is evidence for strong lines due to [ArIII] $\lambda$8.991, [SIV] $\lambda$10.51, [NeIII] $\lambda$15.56, [SIII] $\lambda$18.713, [OIV] $\lambda$25.87, [SIII] $\lambda$33.482, [NeIII] $\lambda$36.0, [OIII] $\lambda$51.816 and [NIII] $\lambda$57.317 $\mu$m. Fluxes at $\lambda$ > 15 $\mu$m are dominated by a broad dust continuum with black body (BB) temperature of ~124 K. Where grain emissivity varies as $\varepsilon \propto \lambda^{-\beta}$, on the other hand, then reasonable fits may be obtained for temperatures of ~95 ($\beta$ = 1), and ~80 K ($\beta$ = 2). The main deviation from these trends occurs over the range 27.5 < $\lambda/\mu$m < 44, where there is an excess attributable to silicate emission bands.

The nature of this excess is more clearly seen in Fig. 3, where we have removed the underlying (smoother dust) continuum using the BB fit (i.e. assuming emissivity exponents $\beta$ = 0), and smoothed the results using an eight-point moving average. Closely similar excesses are also obtained for other values of β. We have attempted to fit the trends using various temperature weighted silicate emission functions, including Forsterite ($Mg_2SiO_4$) and clino- and ortho-enstatite ($MgSiO_3$) (see e.g. the transmission spectra of Jäger et al. (1998) and Koike & Shibai (1998), and the general discussion of silicate emission provided by Molster & Kemper (2005)). The results are only partially satisfactory, and are not shown here. It is plain however that the primary peaks for these materials can be identified within the spectrum.

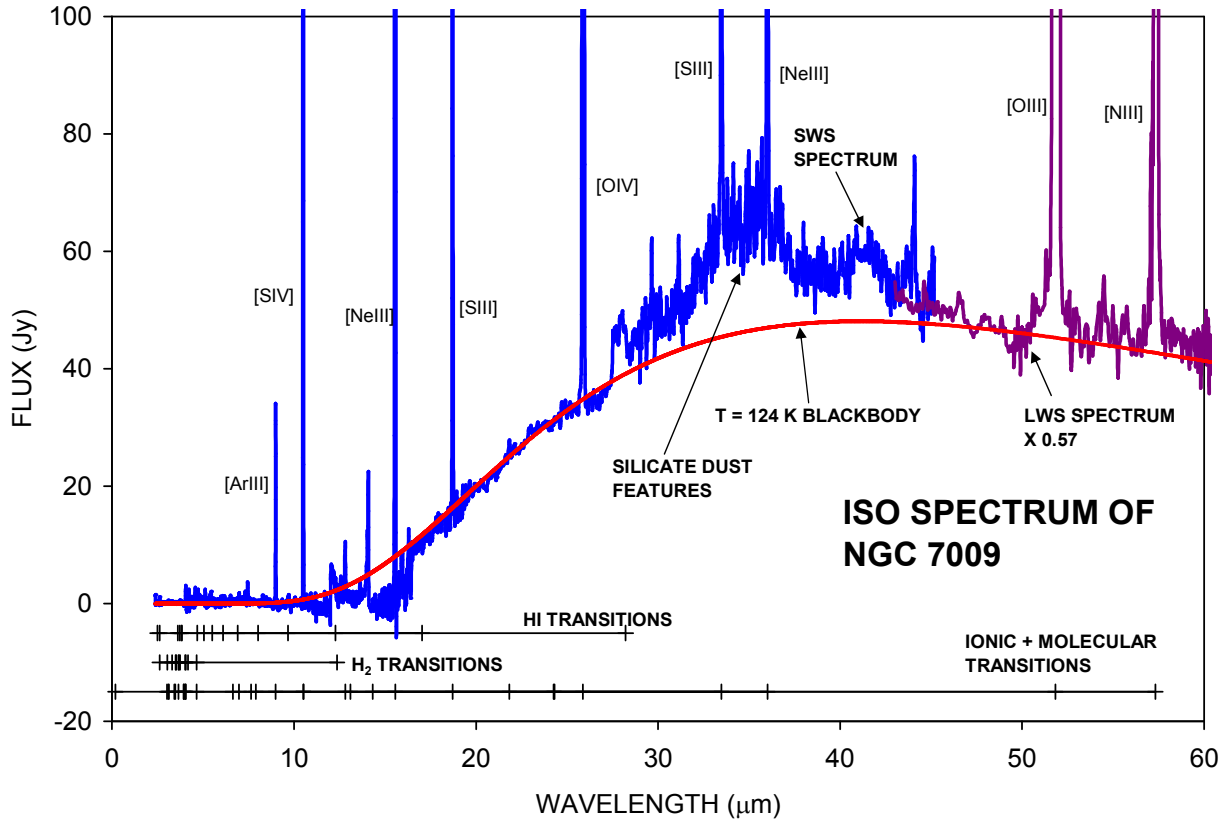

**Figure 2:** Combined ISO spectrum for the interior portions of NGC 7009, where we have included SWS spectra between 2.38 and 45.3 μm, and the LSW spectrum for λ = 45 -198.8 μm. The LWS spectrum has been adjusted by a factor 1.76 to fit the shorter wave LWS results, suggesting that the broad continuum component is appreciably extended. We have also identified certain of the primary emission lines; fitted a representative 124 K BB spectrum to the dust emission component; and noted the presence of a strong excess between 27.5 and 44 μm attributable to crystalline silicate emission.

It would therefore appear that the MIR spectrum is consistent with ionic abundances in suggesting O/C ratios which are greater than unity.

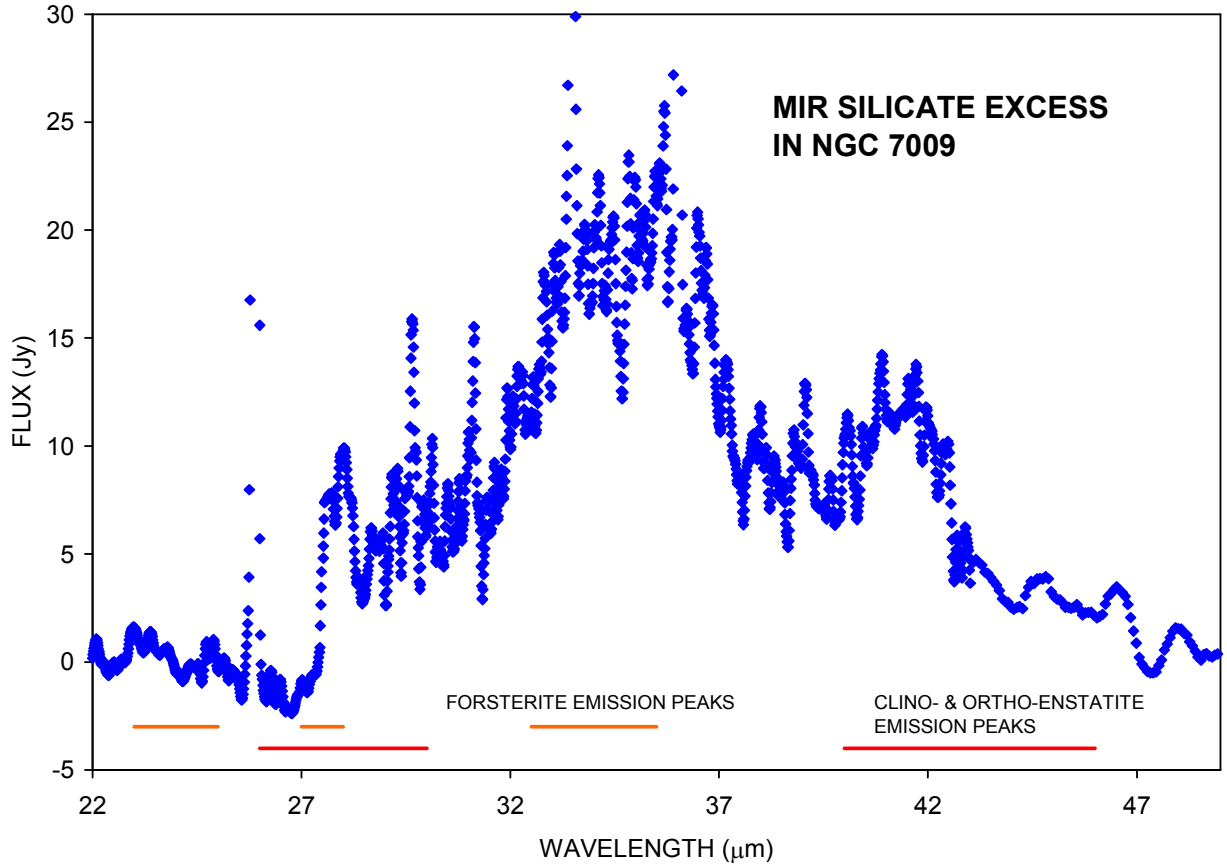

**Figure 3:** The MIR silicate excess in NGC 7009, obtained by subtracting a BB continuum from the broader dust excess. The lower bars indicate the primary emission bands associated with clino- and ortho-enstatite ($MgSiO_3$) and Forsterite ($Mg_2SiO_4$).

An expanded version of the shorter wave ISO and Spitzer spectra is illustrated in Fig. 4, together with filter profiles for the 3.6 μm (blue), 4.5 μm (turquoise), 5.8 μm (green) and 8.0 μm (red) IRAC bands. Both of the spectra were taken at the centre of the source, although with markedly differing apertures. The ISO spectra were acquired using the SWS and LWS (see Sect. 2), with SWS aperture sizes varying from 14x20 arcsec$^2$ between 2.38 and 12.1 μm, 14x27 arcsec$^2$ between 12 and 29 μm, and 20x22 arcsec$^2$ between 29 and 45.3 μm. By contrast, the LWS beam is elliptical as a consequence of the secondary mirror supports, and the inclination of the LSW mirror M2 (Clegg et al. 1996). The FWHM also appears to vary very slightly with wavelength, although its mean value is of the order of ~70 arcsec (see e.g. Polehampton 2002). We have reduced the LWS spectrum by a factor 1.76 to match the LWS spectra close to 45 μm, suggesting that cool dust emission is appreciably extended compared to the bright interior shell (~ 22.5x31 arcsec$^2$). The Spitzer spectrum, by contrast, was taken using the Short-High (SH) module with slit dimensions 11.3x4.7 arcsec$^2$, a resolution of ~600, and a spectral range of between 9.9 and 19.6 $\mu$m. It has been necessary to increase this latter spectrum by a factor $\cong$ 30.7 to fit the corresponding ISO results, as illustrated in Fig. 4. When this adjustment is made, however, then it is clear that both sets of results are similar.

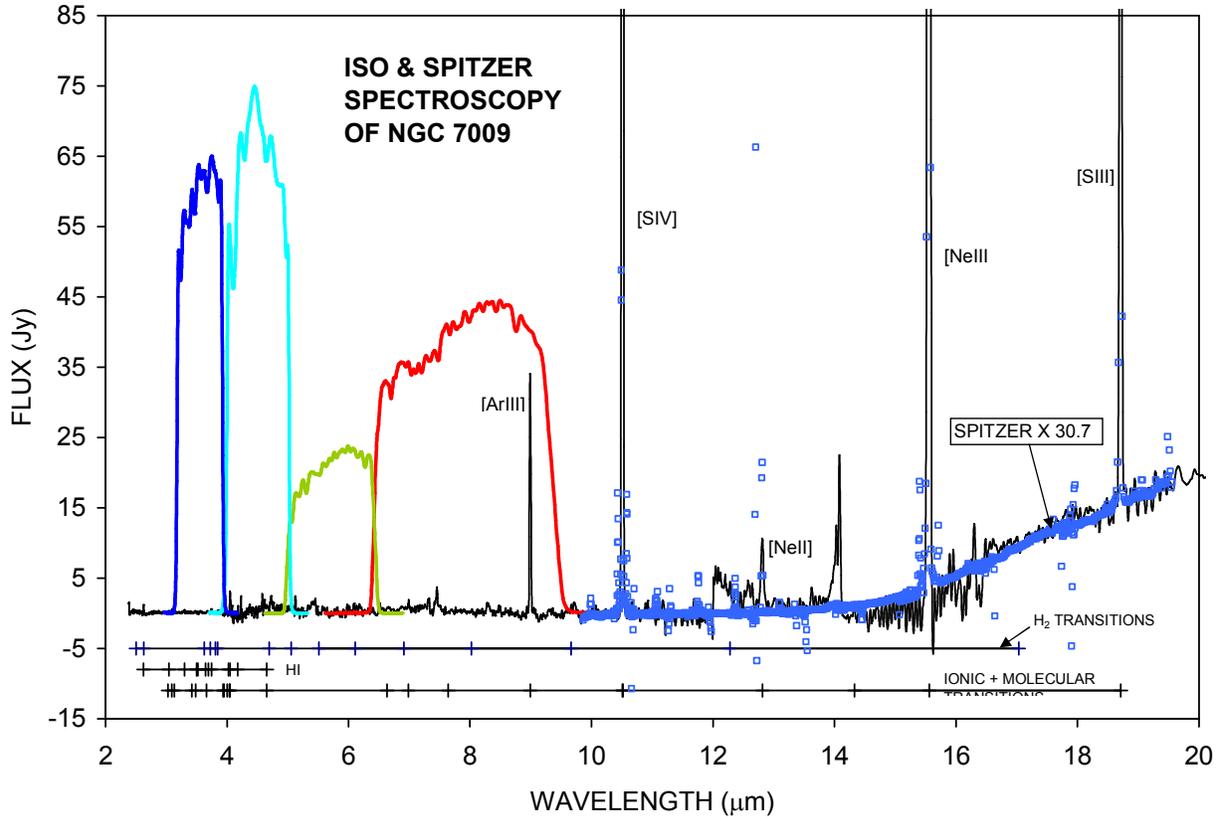

**Figure 4:** The short wavelength regime of the ISO spectrum for NGC 7009, where we have also fitted a Spitzer spectrum for wavelengths 9.9 < $\lambda$ < 19.6 $\mu$m, and indicated the filter profiles for the Infrared Array Camera (IRAC). Note how the IRAC emission at 8.0 $\mu$m (red filter profile) is dominated by the [ArIII] $\lambda$8.991 $\mu$m transition.

It is finally worth noting that although longer wave dust continuum levels are certainly large, their contribution at $\lambda$ < 14 $\mu$m is relatively small. The dominant line in the 8.0 $\mu$m band is [ArIII] $\lambda$8.991 $\mu$m, whilst there is also evidence for transitions such as [SIV] $\lambda$10.51 $\mu$m, He I $\lambda$4.049 $\mu$m and Br$\alpha$ $\lambda$4.052. The variations in morphology in the differing IRAC bands (Sect. 4) can therefore be attributed to changes in the contributions of differing ionic transitions.

Finally, we are unaware of any detection of CO or $H_2$ transitions within NGC 7009 (see e.g. earlier summaries of PNe observations by Huggins et al. (1996) and Kastner et al. (1996)) and this is broadly as might be expected. Central star temperatures are high, and levels of shell excitation are appreciable – even increasing, as we shall note in Sect. 5.2, to larger distances from the central star. It is therefore hardly surprising that we detect little if any evidence for $H_2$ transitions in the present ISO/Spitzer spectra. Whilst there is certainly a peak at the expected position of the v = 0-0 S(13) $\lambda 3.847\,\mu m$ transition of $H_2$, this is very much an exception. The absence of any of the other v = 0-0 or v = 1-0 transitions suggests that this peak is likely to be an anomaly, or attributable to other ionic or neutral species.

## 4. Mid Infrared IRAC Imaging with the SST

An MIR image of NGC 7009 is illustrated in Fig. 1, where we have represented $3.6\,\mu m$ emission as blue, $4.5\,\mu m$ emission as green, and $8.0\,\mu m$ fluxes as red. By contrast, the lower panel shows a corresponding HST image in which HeII $\lambda 4686$ corresponds to blue, [OIII] $\lambda 5007$ to green, and [NII] $\lambda 6583$ is represented by red. It is clear that despite their relatively small dimensions, both ansae and jets are well detected in the MIR, where they appear to be particularly strong at $4.5\,\mu m$ and have a predominantly green colouration. Given the low excitation characteristics of these features at visual wavelengths (see e.g. Sect. 5.2), one might suppose that the MIR emission derives from lower excitation transitions as well. However, although there are certainly several transitions within the $4.5\,\mu m$ band which may contribute to these features (Fig. 4), most of them are difficult to identify. The strongest identifiable transition is that of Br$\alpha$ at $\lambda 4.052\,\mu m$, although we also note that various v = 1-0 CO and v = 0-0 $H_2$ transitions occur in the range $4.649$-$4.694\,\mu m$ - lines which might contribute to a possible shallow excess noted in Fig. 4. We are unable to identify any particular peaks associated with these transitions, however, and it seems improbable that CO and $H_2$ can be important for the source as a whole (see our comments in Sect. 3). This therefore suggests that the ansae/jets appear green because of the absence of higher excitation transitions, with much of the emission deriving from Br$\alpha$, or that shocked or fluorescently excited transitions of $H_2$ are important, but have not been picked up in the ISO spectroscopy – the LWS apertures exclude these outer regions of the source.

The 8.0 μm emission appears to be more extended, and forms an extended halo about the elliptical central shell. This variation in structure is also noted in the contour mapping (Fig. 5), where the shell appears to become larger and more circular towards longer MIR wavelengths. The lowest contours are set at 3σ background noise levels or greater, whilst the flux $E_n$ corresponding to contour $n$ is given by $E_n = A10^{(n-1)B}$ MJy sr$^{-1}$, where $n = 1$ corresponds to the lowest (i.e. the outermost) level. Values of A & B are indicated in the figure caption.

Finally, we have presented profiles along the major axis of the source in Fig. 6 (upper panel), passing through the position of the central star and the two external ansae. Background emission is present in all of the bands, but appears to be particularly strong at 5.8 and 8.0 μm. We have corrected for these by subtracting constant background fluxes, as well as lineal gradients of the order of 5 10$^{-4}$ MJy/sr/pix.

Several aspects of these trends are of interest. In the first place, it is clear that fluxes between 3.6 and 4.5 μm are not too greatly different, in keeping with what might have been anticipated from the ISO spectroscopy illustrated in Figs. 2 & 4. Similarly, it is noted that 4.5 μm emission is somewhat stronger within the nucleus; the 3.6 μm emission appears weaker at all major axis offsets; and the 5.8 and 8.0 μm emission is stronger at relative positions (RPs) of > 20 arcsec. Uncertainties in flux calibration may be sufficient to account for certain of these disparities, however. The flux corrections described in Sect. 4 of the IRAC Instrument handbook imply that surface brightnesses need to be modified in extended sources, with maximum changes being of the order ~0.91 at 3.6 μm, 0.94 at 4.5 μm, 0.66-0.73 at 5.8 μm, and 0.74 at 8.0 μm; although the precise values of these corrections also depend on the distribution of source emission. It therefore follows that the central 5.8 μm surface brightness, when appropriately corrected for these effects, may be less that that observed at 3.6 μm. On the other hand, such corrections are incapable of accounting for the strong increase in surface brightnesses noted in the 8.0 μm channel.

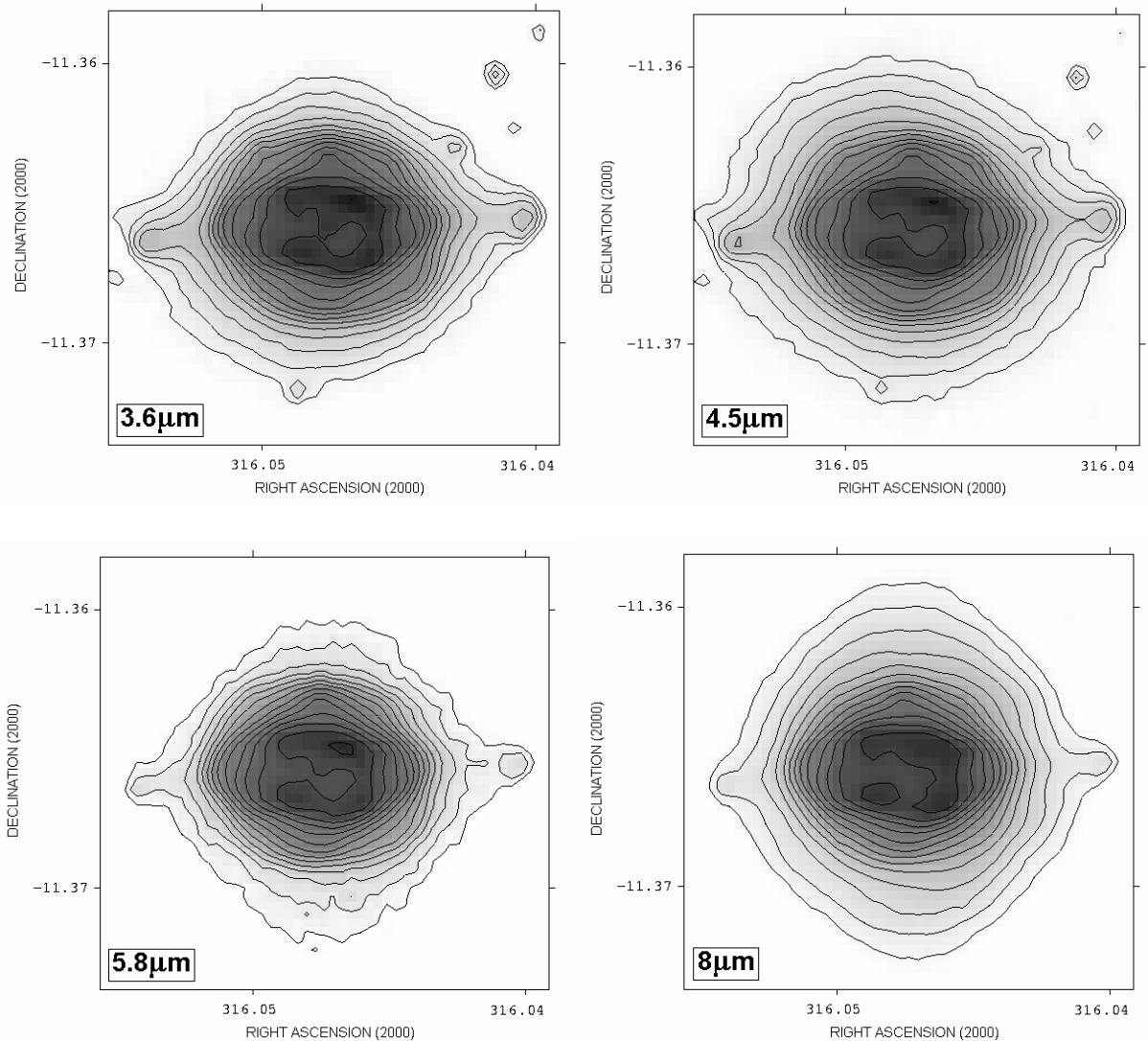

**Figure 5:** Contour mapping of NGC 7009 at mid-infrared wavelengths, where the contour parameters (A, B) are given by (0.2, 0.177) at 3.6 $\mu$m, (0.27, 0.183) at 4.5 $\mu$m, (0.7, 0.134) for 5.8 $\mu$m, and (3.0, 0.133) at 8.0 $\mu$m. Note how the size of the source appears to increase towards longer MIR wavelengths, where the external contours become more circular.

It is also conceivable that scattering in the longer-wave filters may be responsible for some of the more extended emission - although it is unlikely to account for all of the trends noted above. It is worth pointing out for instance that the 8.0 $\mu$m point source function falls-off reasonably quickly, to of the order of ~ 0.1% peak strength at 10 arcsec from the maximum; a variation which is considerably steeper than the changes noted in Fig. 5, or in the profiles illustrated in Fig. 6.

It is finally worth pointing out that the jet and ansae are very clearly discerned in all of the IRAC channels, and lead to secondary peaks at RPs of ± 25 arcsec. Similarly, the 8.0 μm profile is stronger at all positions along the major axis; a trend may presumably be due, at least in part, to the [ArIII] λ8.991 μm transition in Fig. 4.

Ratios between these trends are illustrated in the lower panel of Fig. 6, where it may be noted that the correction factors for the 8.0μm/4.5μm and 5.8μm/4.5μm ratios are likely to be > 0.8, but less than unity. Ignoring this correction therefore has little effect upon our results. It is apparent that the ansae cause dips in the 8.0μm/4.5μm ratio, whilst the corresponding ratios for the jets appear to be enhanced. There is also evidence for an increase in 5.8μm/4.5μm and 8.0μm/4.5μm ratios at larger distances from the nucleus, outside of the regime containing the ansae.

We have finally obtained contour mapping of the 8.0μm/4.5μm and 5.8μm/4.5μm and 3.6μm/4.5μm flux ratios over the source (Fig. 7). This was undertaken by estimating the levels of background emission, removing these from the 3.6, 4.5, 5.8, and 8.0 μm images, and subsequently setting values at < $3\sigma_{rms}$ noise levels to zero. The maps were then ratioed on a pixel-by-pixel basis, and the results contoured using standard IRAF programs. Contour levels are given through $R_n = C10^{(n-1)D}$, where the parameters (*C, D*) are provided in the caption to Fig. 7.

It is again clear that 8.0μm/4.5μm and 5.8μm/4.5μm ratios increase with increasing distance from the nucleus; the emission outside of the central elliptical shell is characterised by significantly higher values of these parameters. This is presumably related to the trends in [OIII]/Hα noted by Bohigas et al. (1994) (see also Sect. 5.2), whereby ratios are found to increase outside of the bright elliptical shell, and suggest higher and increasing excitation. It is therefore apparent that strong higher excitation transitions may be contributing to the longer wave MIR fluxes, and leading to enhanced surface brightnesses compared to those in the 4.5 μm band. One possible culprit (apart from [ArIII] λ8.991 μm) is the transition of [NeVI] at λ7.642 μm. However, the apparent absence of this line in the inner portions of the source suggests that this contribution is likely to be small.

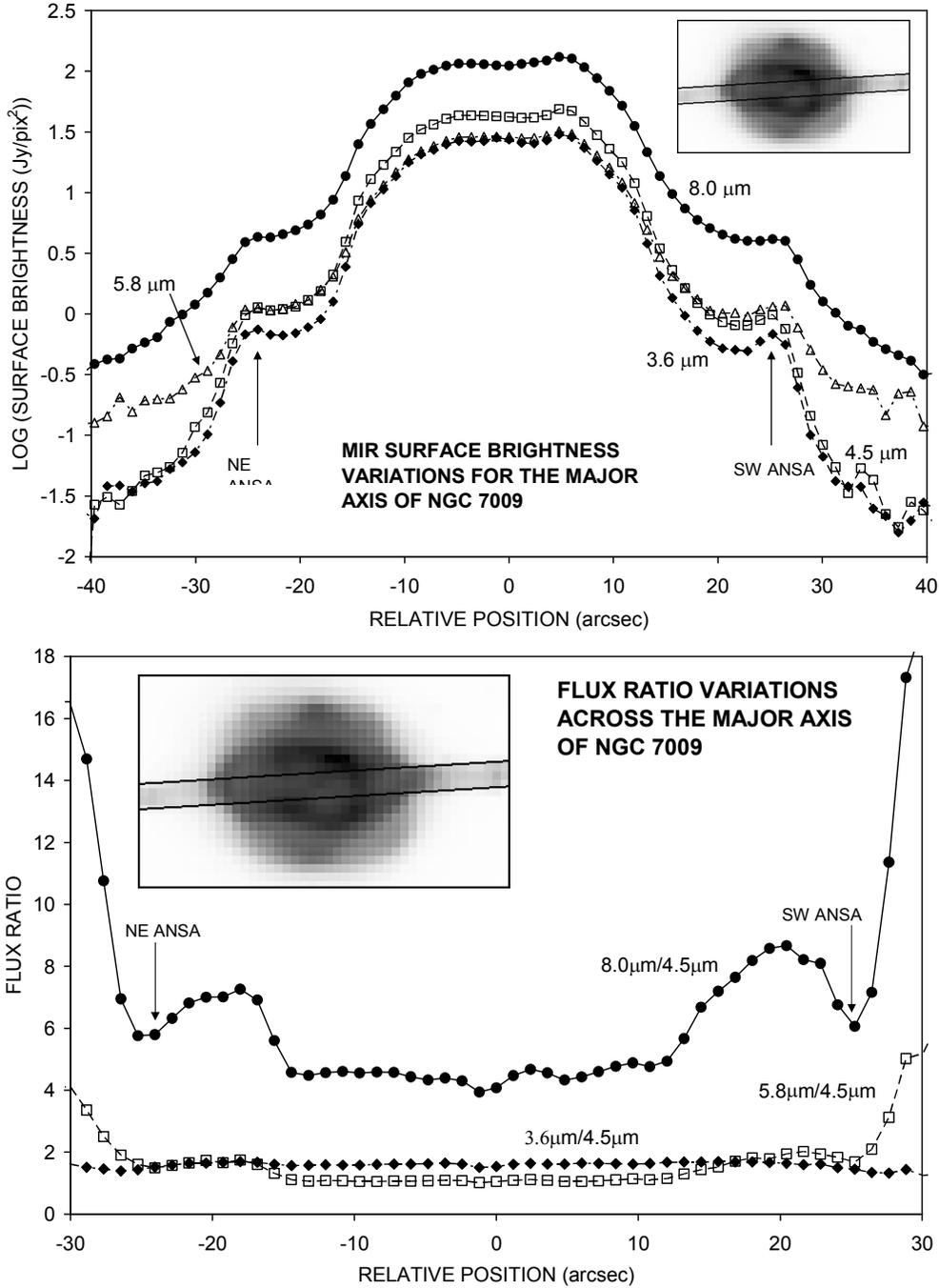

**Figure 6:** MIR surface brightness profiles through the centre of NGC 7009 (upper panel), and the corresponding flux ratios (lower panel). The orientations and widths of the slices are indicated in the inserted images. Among various features of interest, we note the higher levels of surface brightness at 8.0 μm; the presence of more extended emission "skirts" at 4.5 and 8.0 μm, extending to greater than ± 40 arcsec from the centre; the dip in 8.0μm/4.5μm ratios at the positions of the ansae, and the apparent increase in these ratios close to the jet-like features.

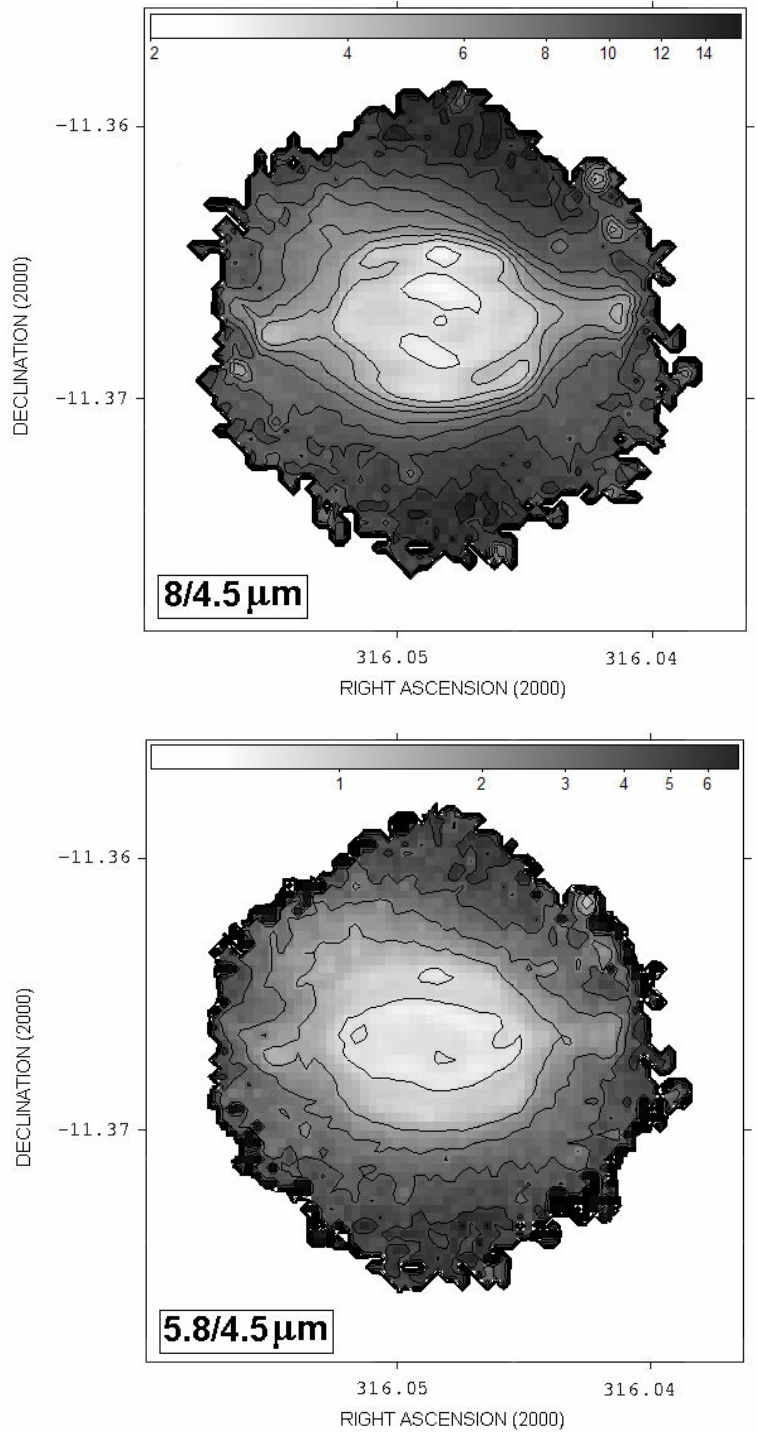

**Figure 7:** Flux ratio mapping for NGC 7009, where we illustrate the variation of 8.0μm/4.5μm ratios (upper panel) and 5.8μm/4.5μm ratios (lower panel). The contour parameters (C, D) are given by (2, 0.1111) in the upper map, and (0.115, 0.1947) in the lower map. Similarly, the ratios corresponding to differing shades of grey are indicated in the upper scale bars. Notice how ratios increase outside of the bright central shell and ansae.

## 5. Long Slit Visual Spectroscopy

Long slit spectra have been obtained for seven PAs across the nucleus of the source, as described in Sect. 2. The slit positions are illustrated in Fig. 8, superimposed upon an HST image taken using filter F658N ([NII] λ6583). We have also indicated six sectors of the shell labelled I through VI, corresponding to the nucleus (I); the bright elliptical inner rim structure (II); the fainter, more spherical enveloping shell (III); the jet-like feature (IV); the ansae (V); and the external halo (VI). Mean spectral line fluxes for each of these sectors are presented in Table 1, where we present line strengths relative to Hβ, uncertainties in the estimates (columns labelled σ), and the calibrated Hβ line flux (in erg cm$^{-2}$ s$^{-1}$). Possible blended lines are indicated in parenthesis in the first column of the Table, whilst total emission for all of the sectors (and slits) is indicated by the letter T, and detailed in columns 3 & 4.

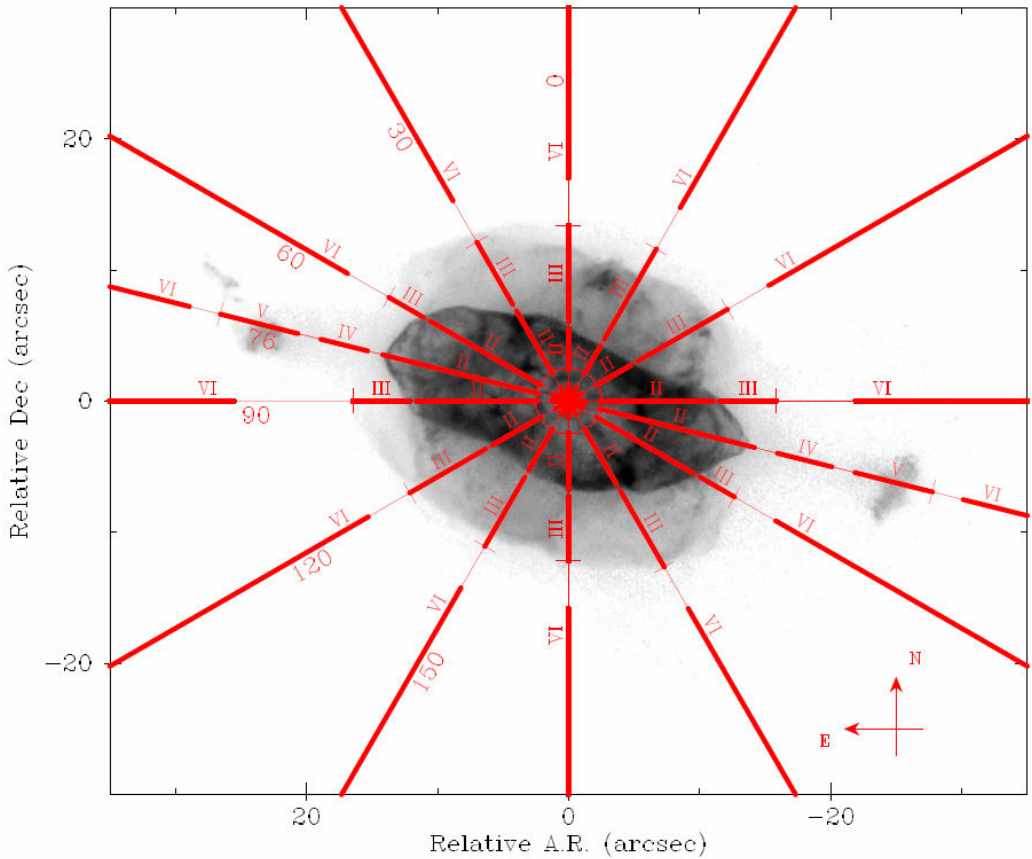

**Figure 8:** The positions of the Boller & Chivens Spectrographic slits upon an HST image of NGC 7009 taken using the F658N ([NII] λ6583) filter. The numbers I through VI correspond to the segments of the shell referred to in Table 1, where I corresponds to the nucleus.

These data will be subject to more detailed modelling in a later paper on this source. For the present, however, we wish to point out several general characteristics of these results, of relevance in interpreting the inner shell, and determining the nature of the ansae and jets.

**5.1 Density and Temperature Structure**

We have obtained profiles of temperature through the source by using the [OIII] ($\lambda$4959+$\lambda$5007)/$\lambda$4363 and [NII] ($\lambda$6548+$\lambda$6548)/$\lambda$5755 line ratios. The results for the major axis (PA = 76°) are illustrated in Fig. 9. It is clear from this that although there is a significant difference (~ 20%) in values outside of the nucleus, values of $T_e$ are, for the most part, invariant throughout the shell. Such a result is consistent with the previous analysis of Tsamis et al. (2008), Rubin et al. (2002), Sabbadin et al. (2004), and Gonçalves et al. (2004, 2005). They are also comparable to the estimates of $T_e$ published by Hyung & Aller (1995a, b), Bohigas et al. (1994), and Balick et al. (1994). It is particularly interesting to compare our present results with the VLT FLAMES Argus data of Tsamis et al. (2008), who obtained maps of density and temperature for one side of the interior shell. Their mapping of $T_e$([OIII]) includes the central 11.5 arcsec segment of the profile in Fig. 9, for which they find values ranging from ~0.98 $10^4$ K at the outer limits of the map (i.e. at the approximate PAs = ± 5.5 arcsec), rising to a peak of ~1.05 $10^4$ K at the centre of the source. This is identical to the variation in $T_e$(OIII) noted in Fig. 9. The authors also note small-scale spaxel-to-spaxel variations in $T_e$, for which there is also marginal evidence in our present results.

Densities, by contrast, have been determined using the Lick five level atomic program of Shaw & Dufour (1995) in an interactive version due to R.J. Dufour, and show more significant variations with respect to axial position. Values of $n_e$ appear closely similar whether one uses $T_e$([NII]) or $T_e$([OIII]), and show peaks at ≈ +9 and -13 arcsec from the centre, close to the limits of the elliptical inner shell. These peak densities are also close to the position of the internal FLIERs described in Sect. 1.

Outside of this regime, however, in the region of the external jets and ansae, it is clear that densities $n_e$ fall to significantly lower values (of the order of 900-2600 cm$^{-3}$); a variation which is very closely similar to that noted in several previous studies. Thus, the trends and densities are comparable to those of Gonçalves et al. (2003), excepting that the latter authors find larger internal densities for -8 arcsec < RP < 3 arcsec. A similar fall-off in densities from the edge of the internal shell (RPs close to ~10.5 arcsec) has also been found by Balick et al. (1994), who determine a variation 4-5 $10^3$ – $10^3$ cm$^{-3}$ comparable to that observed here.

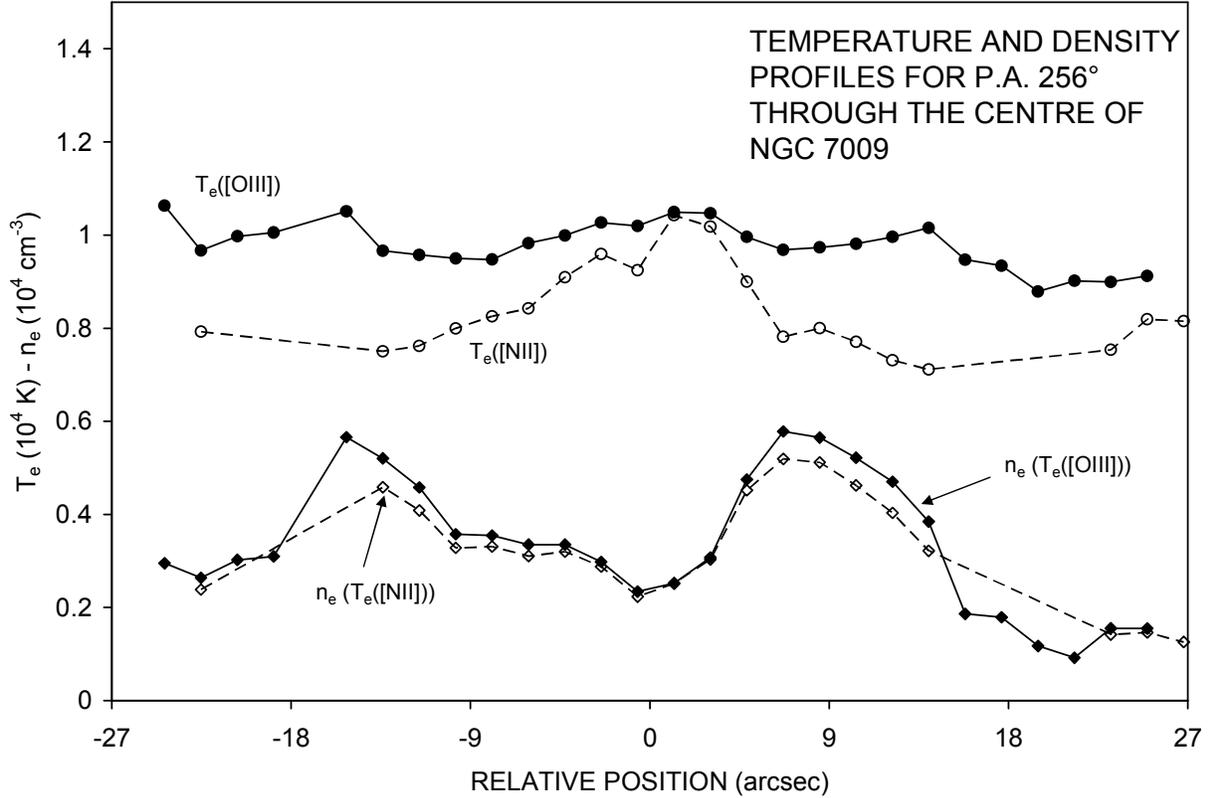

**Figure 9:** Trends in temperature and density along the major axis of NGC 7009, where $T_e([NII])$ is determined using the $(\lambda 6548+\lambda 6548)/\lambda 5755$ line ratios, $T_e([OIII])$ makes use of the $(\lambda 4959+\lambda 5007)/\lambda 4363$ ratios, and densities $n_e$ are derived using the [SII] $\lambda 6717/\lambda 6731$ doublet. There appears to be some slight difference in our estimates of $T_e([OIII])$ and $T_e([NII])$, with the former being ~ 20 % larger outside of the central portions of the source. Similarly, it would appear that densities peak close to the edge of the bright elliptical shell, where $n_e$ ~ 5.8 $10^3$ cm$^{-3}$ and RP ~ -13/+9 arcsec, and decline steeply at the positions of the jets and ansae (where $n_e$~ 0.9-2.6 $10^3$ cm$^{-3}$).

Finally, the detailed mapping of Tsamnis et al. (2008) appears to show appreciable variations over a range of 4-5 $10^3$ cm$^{-3}$, and over spaxel distances of ~0.5 arcsec. If these are real, they appear to be largely smoothed out in our lower resolution data. The Tsamis et al. mapping also shows mean values of $n_e$(Ar IV) which are ~20 % less than our present $n_e$([SII]) results. This may reflect physical differences between the two ionic regimes, such as may be responsible for differences in the [Cl IIII] and [SII] densities determined by Corradi et al. (2004).

We have finally taken this analysis a little bit further in Fig. 10, where we have determined density and temperature profiles along each of the PAs, and used a process of annular interpolation to create two-dimensional mapping of the parameters. Although the resolution of these maps is relatively low, they enable us to gain a global perspective concerning changes in the parameters. It is clear from this that temperatures do appear to change throughout the shell, although the range of variation is modest and of order ~8000-$10^4$ K. Similarly, although there appear to be compact regions where densities approach ~$10^4$ cm$^{-3}$ (as also found for instance by Lame & Pogge 1996), the mean density of the interior shell is closer to ~5 $10^3$ cm$^{-3}$.

It should be noted that we have used an extinction c ~ 0.15 in calculating these values, consistent with the broad range of values previously published for this parameter (see e.g. Tsamis et al. (2008) (0.18); Sabbadin et al. 2004 (0.15); Perinotto & Benvenuti 1981 (0.15); Tylenda et al. 1992 (0.39-0.56); Kingsburgh & Barlow 1994 (0.09); Bohigas et al. 1994 (0.05-0.28); Liu et al. 1995 (0.20); Hyung & Aller 1995a,b (0.09); Lame & Pogge 1996 (0.24); Ciardullo et al. 1999 (0.14); Luo et al. 2001 (0.07); Rubin et al. 2002 (0.10); Gonçalves et al. 2003 (0.16)). Bohigas et al. (1994) find little evidence for variations in extinction across the shell, although Tsamis et al. (2008) suggest an rms deviation of 0.111, with slightly higher values (c ~ 0.15) occurring near the bright interior rim.

**5.2 Excitation Structure**

Interpolation between the profiles can also be used to provide two dimensional mapping of excitation, and examples of this are illustrated in Fig. 11. It should again be emphasised that although resolutions are modest, they are sufficient to gain an accurate global perspective of the trends, whilst the range of transitions is larger than available through narrow band imaging. Such mapping is possible, in principal, for all of the lines in Table 1, although such an analysis only makes sense where S/Ns are reasonably high.

The maps show the increase in excitation towards the outer parts of the source previously noted by Bohigas et al. (1994) (viz. the map of [OIII]/H$\beta$ in the upper left-hand panel, and of [ArIV]/H$\beta$ in the lower right-hand panel), as well as the increases in lower excitation emission in the jets, ansae, and interior FLIERs.

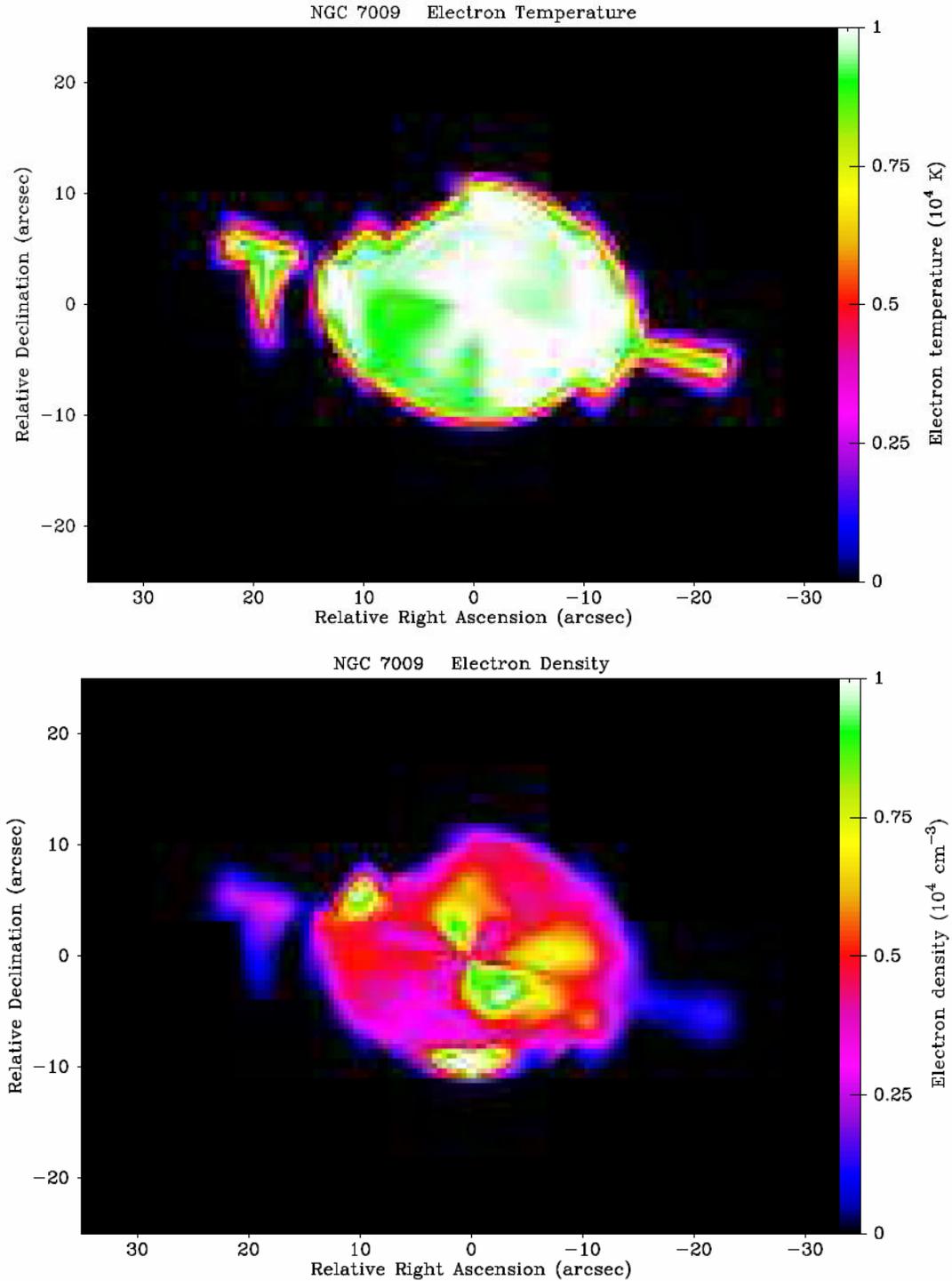

**Figure 10:** Two dimensional maps of temperature (upper panel) and density (lower panel), where we have determined values of $n_e$ and $T_e$ along the seven slit PAs, and evaluated the 2D variations through a process of annular interpolation. Notice how temperatures appear to vary throughout the source, although the range of values is modest (~ $8 \times 10^3 - 10^4$ K). Similarly, although interior shell densities are of the order of $5 \times 10^3$ cm$^{-3}$, there are variations from ≈ $10^3$ cm$^{-3}$ (at the ansae) through to as high as ≈ $10^4$ cm$^{-3}$ (in the interior of the source).

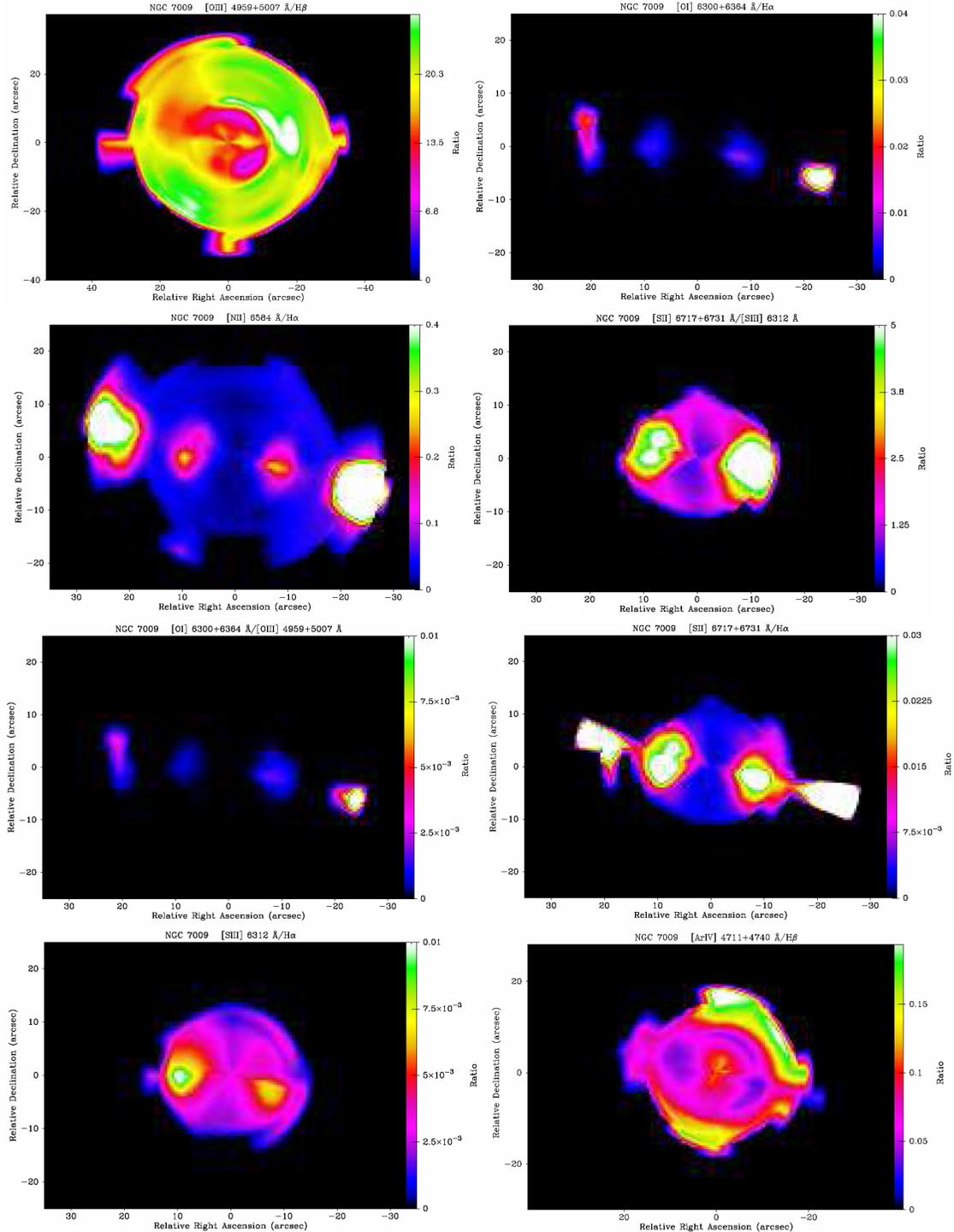

**Figure 11:** Excitation mapping of NGC 7009, where we show interpolated distributions for a variety of line ratios. Note the tendency for higher excitation transitions to become relatively stronger in the outer portions of the shell (i.e. in the sector VI indicated in Fig. 8). The lower excitation transitions, by contrast, peak at the ansae, the jets, and at the major axis limits of the bright elliptical shell, where low-excitation FLIERs are located.

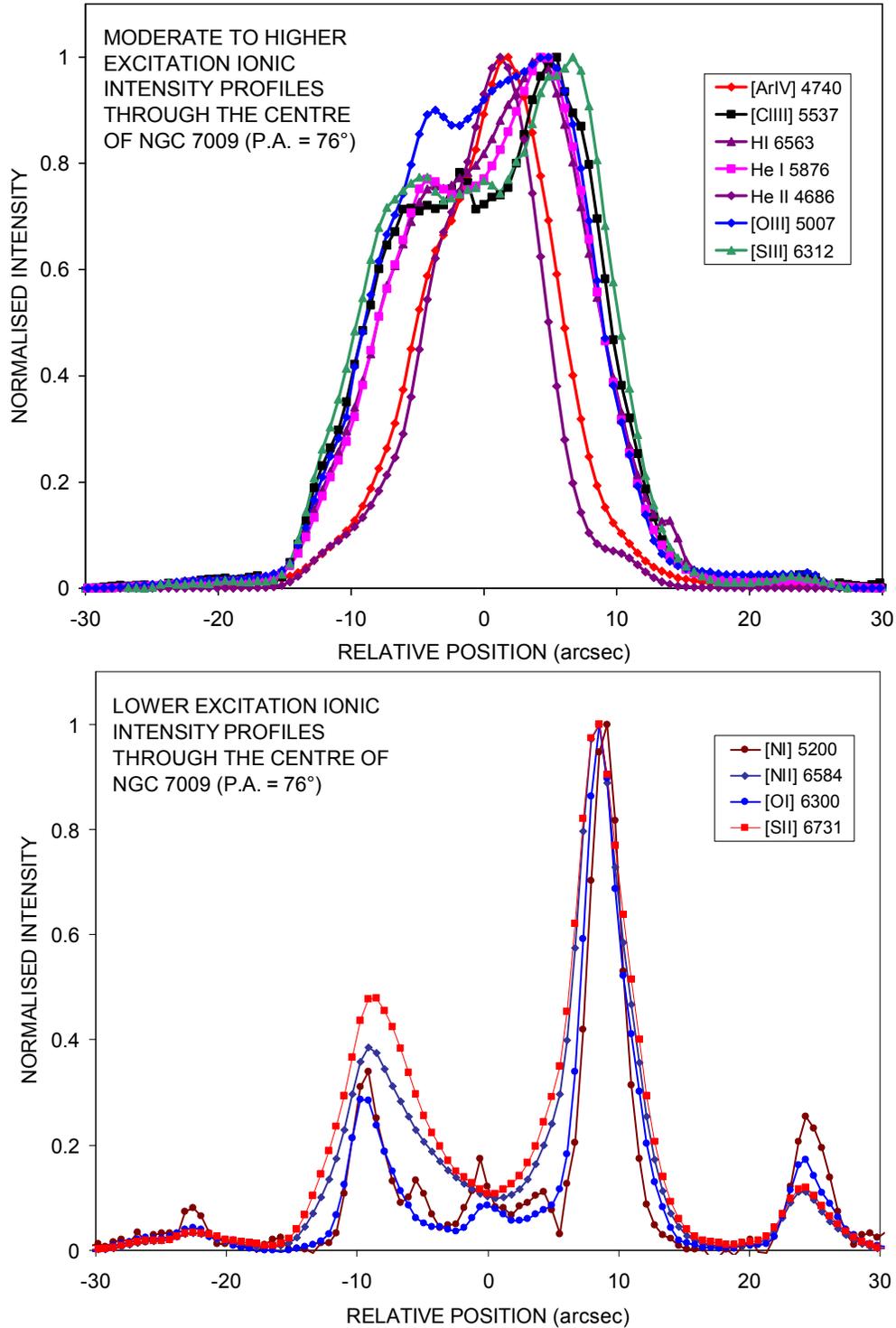

**Figure 12:** Major axis spatial profiles for higher excitation lines (upper panel) and lower excitation transitions (lower panel). The differences between the profiles are broadly as would be expected, and show the concentration of higher excitation emission towards the nucleus, and of lower excitation emission within the interior FLIERs and external ansae.

Differences between lower and higher excitation trends are also apparent in Figs. 12 and 13, where we show normalised profiles for ionic line intensities and ratios. It is particularly interesting to note how [NI]/H$\alpha$ appears to have a narrower peak at RP= +25.6 arcsec compared to other low excitation transitions – and that this peak occurs at smaller distances from the nucleus. This is a somewhat surprising result, and inconsistent with what might be expected for ionisation stratification or shock excitation. It is possible that we are dealing, however, with denser regions of neutral gas (knots, shocks fronts or the like), surrounded by an ionised regime which extends to larger distances from the nucleus.

It is also noteworthy, in the case of the ansae, that although there are small differences in the FWHM of the [NII]/H$\alpha$ and [SII]/H$\alpha$ profiles in Fig. 13, and between the corresponding [NII] and [SII], and [NI] and [OI] profiles in Fig. 12, they are for the most part closely comparable. The present results therefore confirm the lower S/N trends of Gonçalves et al. (2003), and offer support for the suggestion that the N abundances of these features are similar to those in other portions of the shell – a gross increase in N/H might have been expected to lead to differences in the profiles.

The higher excitation transitions show a broader range of variations with respect to H$\alpha$. Thus, the upper panel of Fig. 13 shows that whilst HeII and [ArIV] are strongly concentrated towards the centre, [Cl III] and [SIII] peak at the limits of the elliptical shell (at RP ≈ ± 10 arcsec), in the regime of the interior FLIERs described in Sect. 1. The apparent increase in [OIII]/H$\alpha$ towards the jets and ansae may be no more than part of the generalised increase along all radial directions, noted in the mapping in Fig. 11.

Finally, we have determined the major axis variations of H$\alpha$/[NII] and [SII]$\lambda$6717/$\lambda$6731 as a function of H$\alpha$/[SII] (see Fig. 14). We have also indicated the regions of line ratio expected for supernovae remnants (SNRs) and HII regions (taken from Sabbadin, Minello, & Bianchini 1977), as well as those for PNe (taken from Riesgo & Lopez 2006). It is apparent that there are systematic trends in these parameters with respect to position within the source. Values of Hα/[NII] and Hα/[SII] are much larger at the centre, for instance, than they are towards the edges, where the ansae and jets are located. Nevertheless, all of the values are consistent with those observed in other PNe, and inconsistent with the shock excited ratios found in SNR.

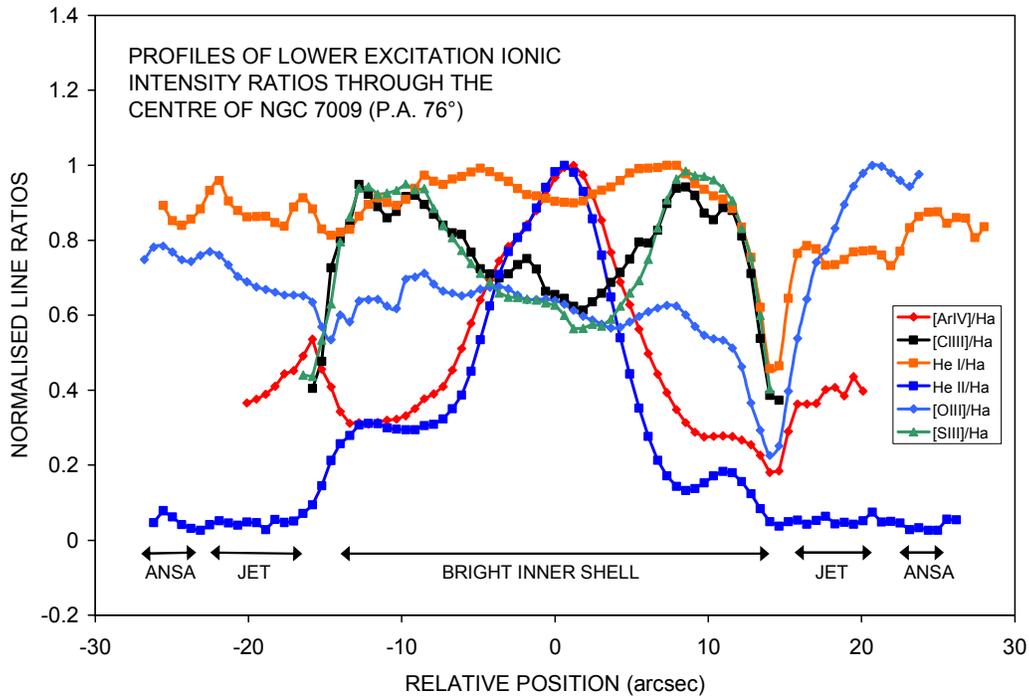

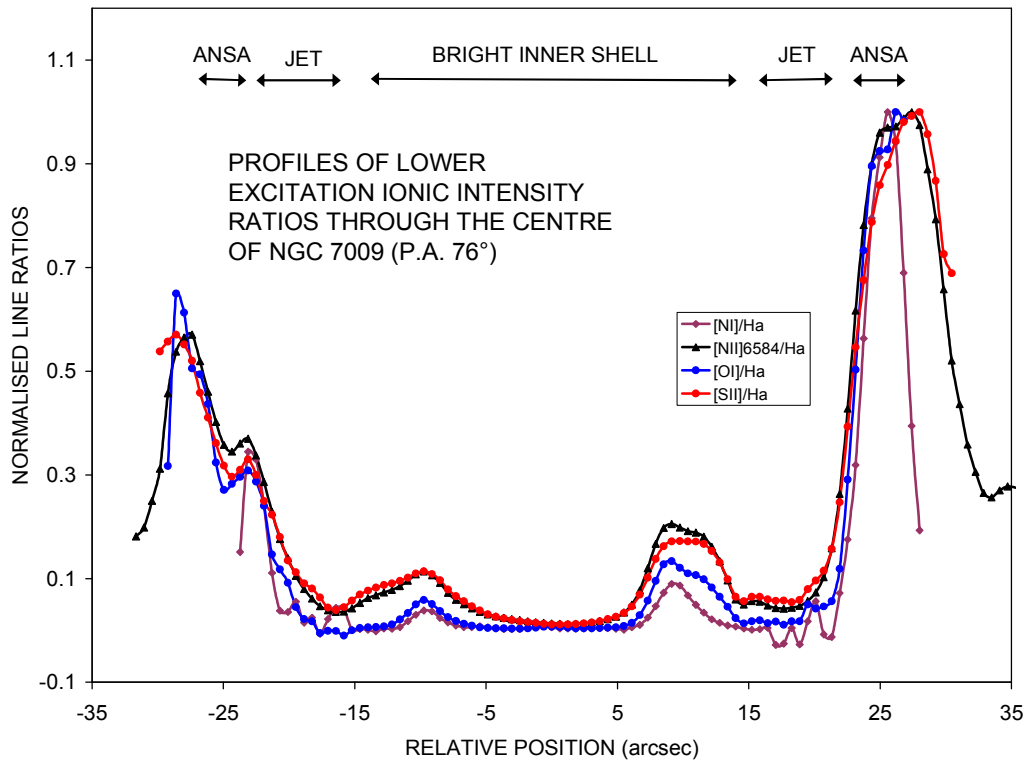

**Figure 13:** As for Fig. 12, but for ratios of selected ionic transitions with Hα. The most centrally concentrated transitions correspond to those of [ArIV] and HeII, whilst certain other ions ([SIII], [ClIII]) are stronger in the inner FLIERs (RP ≈ ± 10.5 arcsec). All of the lower excitation transitions, on the other hand, show stronger relative emission in the ansae, with the profiles for [NI]/Hα appearing to be unusually narrow.

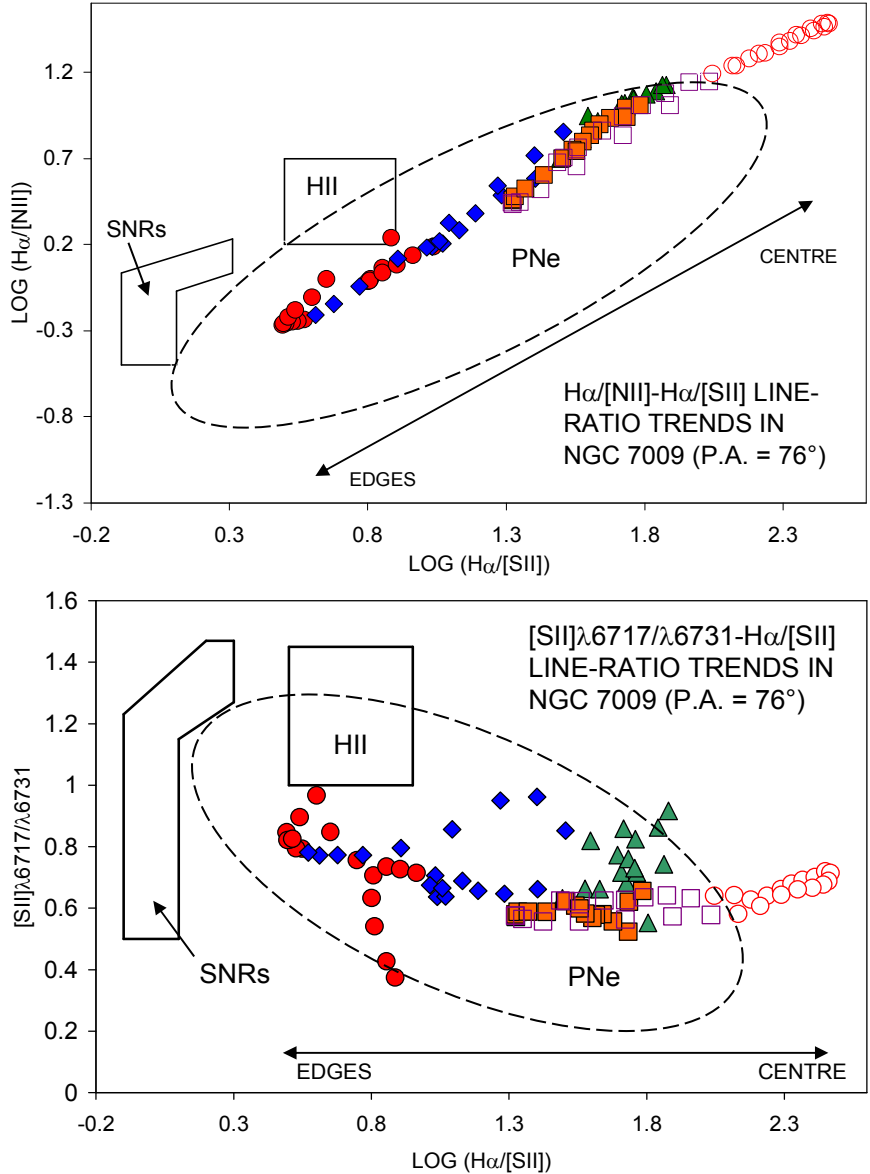

**Figure 14:** The distribution of the major axis ratios Hα/[SII] against Hα/[NII] (upper panel) and [SII] λ6717/λ6731 (lower panel). In both cases, open circles correspond to a range of distances Δr = 0-5 arcsec from the centre; open squares are for Δr = 5-10 arcsec; filled squares are for Δr = 10-15 arcsec; filled triangles correspond to Δr = 15-20 arcsec; filled diamonds are for Δr = 20-25 arcsec; and filled circles are for Δr = 25-30 arcsec. We also show the ratios expected for supernovae remnants and HII regions (taken from Sabbadin, Minello, & Bianchini 1977), and for planetary nebulae (taken from Riesgo & Lopez 2006). It will be noted that ratios towards the edges of the source (close to the ansae) are located to the left of the figures, whilst those corresponding to the centre of the source are positioned to the right. In neither case, however, is there any indication for the kind of shock excited ratios found in SNRs. Ratios are, taken as a whole, typical of those for (radiatively excited) PNe.

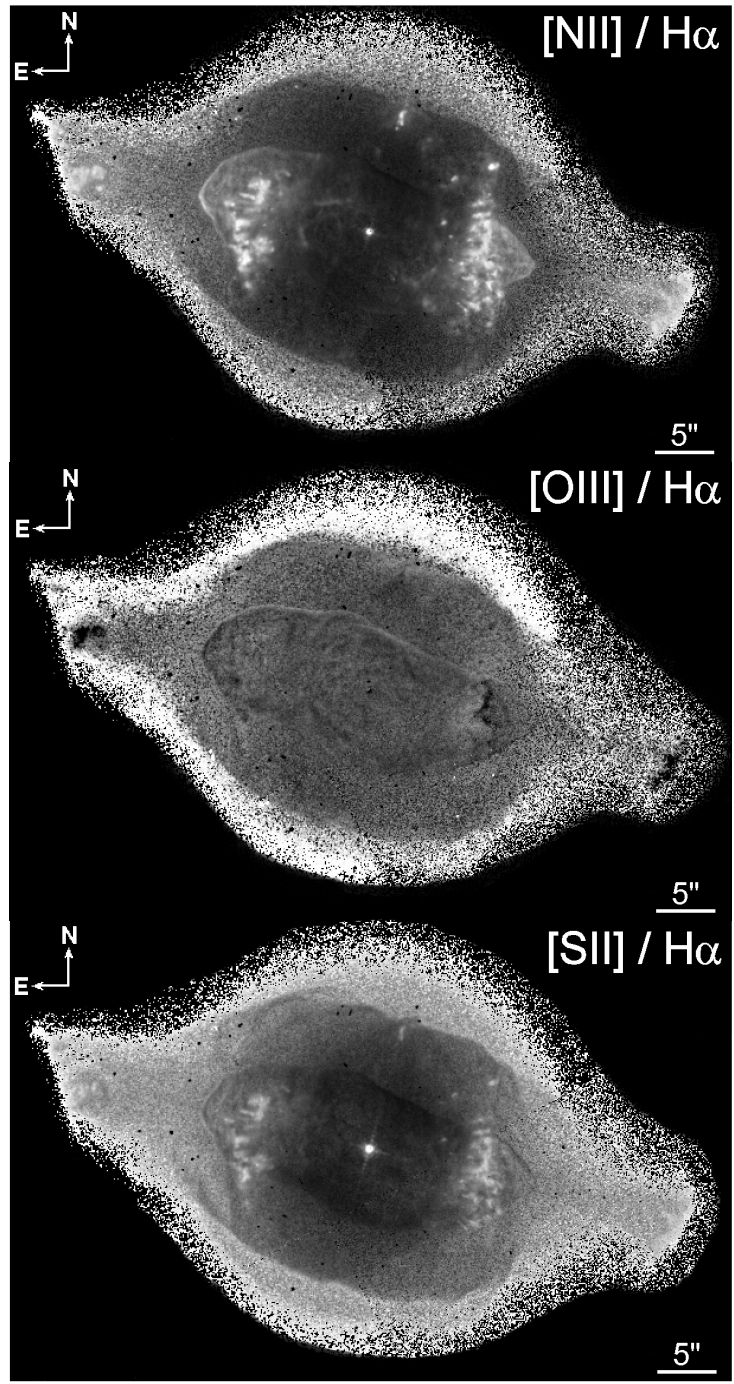

**Figure 15:** Line ratio imaging for [NII]/Hα, [OIII]/Hα and [SII]/Hα, where the lowest values of ratio are indicated using darker shades of grey. Among the several interesting features which are visible in these images, we take note of the sharply defined, thin emission rim at the edge of the interior elliptical shell; the increase in ratios within the outer spherical halo regime; and the prominence of the low excitation ansae and FLIERs, leading to enhanced ratios in the [NII]/Hα and [SII]/Hα maps, and reduced ratios for [OIII]/Hα.

Such a result is also in agreement with other studies of this source, as noted in Sect. 1, and would appear to conflict with the suggestion that the ansae correspond to bow-shock features. We shall suggest a possible resolution of this problem in our analysis of the HST results (Sect. 7).

## 6. Line Ratio Mapping with the HST, and the Detection of Rings and Shocks

Although multi-PA ground-based spectroscopy enables us to determine global trends in emission line ratios (see our discussion in Sect. 5.2), this is not of very great use when investigating finer details of the source structure. We have therefore used HST results to undertaken a ratioing of [SII] $\lambda\lambda$6716+6731, [NII] $\lambda$6583, and [OIII] $\lambda$5007 images with a corresponding exposure in Hα (see Fig. 15). The protocol used in obtaining these maps is similar to that described for the ratioing of the IRAC results (see Sect. 4), although it also requires correcting for differences in filter throughputs $QT$ (where $Q$ is the detector quantum efficiency, and $T$ is filter transmission). Thus, where the image count rate is given by $P$ electrons s$^{-1}$ pixel$^{-1}$, and the wavelength of the transition is $\lambda$ Å, then the surface brightness is given through $I_\nu = P/(4.8\ 10^9 x(QT)x\lambda)$ erg s$^{-1}$ cm$^{-2}$ arcsec$^{-2}$. Further details of the processing of these results, derivation of fluxes, and instrumental properties can be found in the Wide Field and Planetary Camera 2 Instrument Handbook (Biretta 1996), and in Holtzman et al. (1995). Values of $QT$ may be derived from the relevant passband plots, available in Appendix A of the WFPC2 handbook.

Several interesting characteristics are immediately apparent from a cursory inspection of the results. It is apparent for instance that all of the ratios increase from lower values at the centre (where the shades of grey are deeper) towards the circular external shell. The ansae and interior FLIERs also stand out as low excitation condensations in the [OIII]/Hα map, and as enhancements in the corresponding [NII]/Hα and [SII]/Hα images. One of the more fascinating aspects of these present results, however, is the evidence for markedly differing ratios at the limits of the interior ellipsoidal shell. We see evidence for a decrease in [OIII]/H$\alpha$ ratios at the easterly limits of the shell, for instance, and an appreciable increase in [NII]/H$\alpha$. The situation with regard to [SII]/H$\alpha$ appears to be somewhat more complex, and there is evidence for two parallel rims, with increases in the outer rim ratios, and decreases in the inner ratios.

These trends are also seen in Fig. 16, where we show profiles across a narrow sector of the interface. It is clear from this that the [OIII]/H$\alpha$ and [SII]/H$\alpha$ absorptions are displaced by $\sim$ 0.4 arcsec from the [NII]/H$\alpha$ peak, whilst the [SII]/H$\alpha$ enhancement is weak, and not easily discerned in these results. Perusal of Fig. 15, however, suggests that it peaks close to RP $\sim$ -0.2 arcsec; very close to the position of the [NII]/H$\alpha$ peak.

Such sharply defined rims have previously been noted in the present nebula, and in several other PNe as well (Balick et al. 2004; Medina et al. 2009), where they have been attributed to the shock excitation of local nebular material. They are also apparent in various of the transitions mapped by Tsamis et al. (2008) (although the spatial resolution of the latter authors is less). It is therefore possible to imagine that the inner bright elliptical shell is supersonically expanding into the slower exterior shell, and that this is leading to strong local heating at the interface of the shells. Such a hypothesis is consistent with the difference in positions noted between the [OIII] absorption trough, and [SII] and [NII] emission peaks (see e.g. post-shock modelling of Ohtani (1980) and Dopita (1997)), since lower excitation transitions are expected to occur downstream of the shock. The apparent decrease in [SII]/H$\alpha$ at RP $\sim$ 0.4 arcsec may arise because H$\alpha$ peaks at smaller radial distances, and closer to the position of the shock interface (see e.g. Shull & McKee 1979). It would therefore appear that the HST imaging is resolving the post-shock regime, and that the dimensions of the cooling zone are of order $\approx$1 arcsec - a value which corresponds to $\approx$1.3 $10^{16}$ cm at the (uncertain) distance of 0.86 kpc (Fernandez et al. 2004). Such a value is similar to the cooling widths determined from modelling of shocks (see e.g. Shull & McKee (1979) and Dopita (1997), and the analysis of post-shock cooling lengths by Pittard et al. (2005)).

Finally, the apparent decrease in [OIII]/H$\alpha$, and increases in [SII]/H$\alpha$ and [NII]/H$\alpha$ are consistent with what would be expected from modelling of shock transition strengths, as will be noted in our later discussion in Sect. 7 (see also Dopita (1977), Hartigan et al. (1987) and Shull & McKee (1979)).

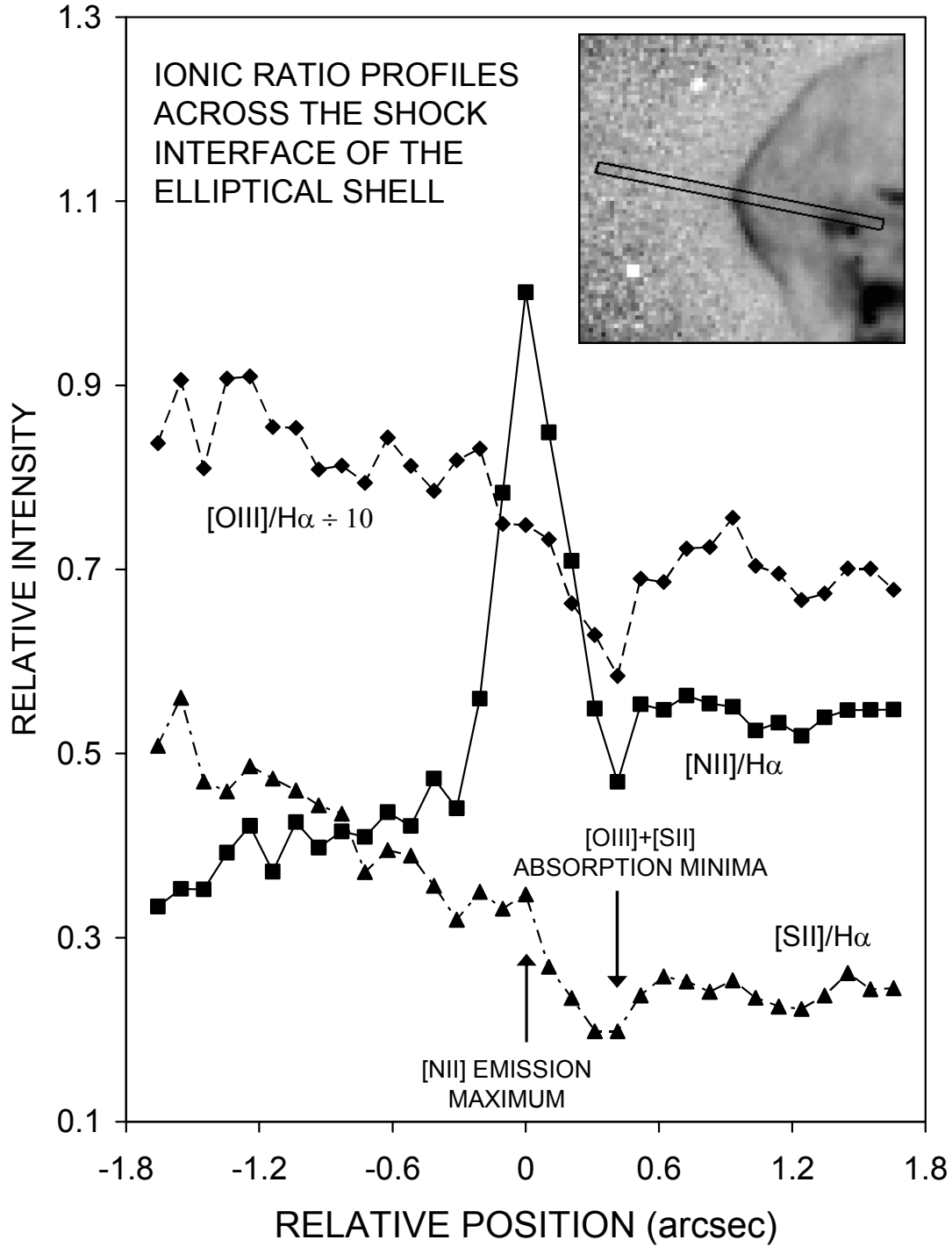

**Figure 16:** Profiles for three ionic line ratios across the easterly limits of the elliptical shell, where the width of the profiles is 2 pixels (corresponding to 0.2 arcsec). Relative profile ratios are correctly specified, whilst the profile for [NII]/H$\alpha$ is normalised to unity. Note that whilst the [NII]/H$\alpha$ profile appears to increase over a narrow range of distances, that for [OIII]/H$\alpha$ decreases, and [SII]/H$\alpha$ has both an increase (peaking at RP $\sim$ -1 arcsec) and decrease (at RP $\sim$ 0.4 arcsec).

Even stronger evidence of shock activity is to be found in deep [OIII] exposures of the source, as indicated in the unsharp masked image illustrated in Fig. 17. This is similar to a previous image of Balick et al. (1998) taken with the 4 m Mayall telescope at the Kitt Peak National Observatory, but shows much finer details than were possible with these previous observations. We again see a region of enhanced emission at the limits of the interior elliptical shell (although it is clear that levels of H$\alpha$ emission must be even greater to explain the decrease in [OIII]/H$\alpha$ ratios noted above). It is also noteworthy that most of the interior shell possesses a mottled and filamentary appearance, perhaps arising from instabilities in the shocks at the limits of the ellipsoidal shell.

A variety of bow-shock type features appear to be associated with the two ansae regimes. Thus, whilst the westerly (right-hand) ansae may represent the primary bow-shock work-face, and is centred on a single emission arc, the easterly ansa is more complex and radially extended, and appears to be related to several differently oriented arcs. The outermost of the condensations at ~31 arcsec from the centre is oriented along a PA of 71°, for instance, whilst the brightest feature at ~25 arcsec from the central star has a PA of close to 82°. Such changes in the orientation of PNe ansae are often attributed to precession of a central collimating engine.

Given that the ansae have velocities of ~ 114 km s$^{-1}$ in the plane of the sky (Fernandez et al. 2004), and that the distance to the source is ~ 0.86 kpc, then the radial difference between the outer and inner features corresponds to a time-scale of 215 yr. The shift in the PAs of ~11° would then imply an angular precession of 5.1 10$^{-2}$ °/yr. If this interpretation is correct, however, then it is of interest to speculate why the westerly ansa shows less evidence for precession. Perhaps the angular width of the ansa is also a consequence of jet precession – or conceivably, it may be possible that there is no precession in this source at all. The complexity of the easterly (left-hand) ansa may simply reflect irregularities in the region of interaction; that the ansae are propagating to larger distances at lower PAs because of lower ambient gas densities.

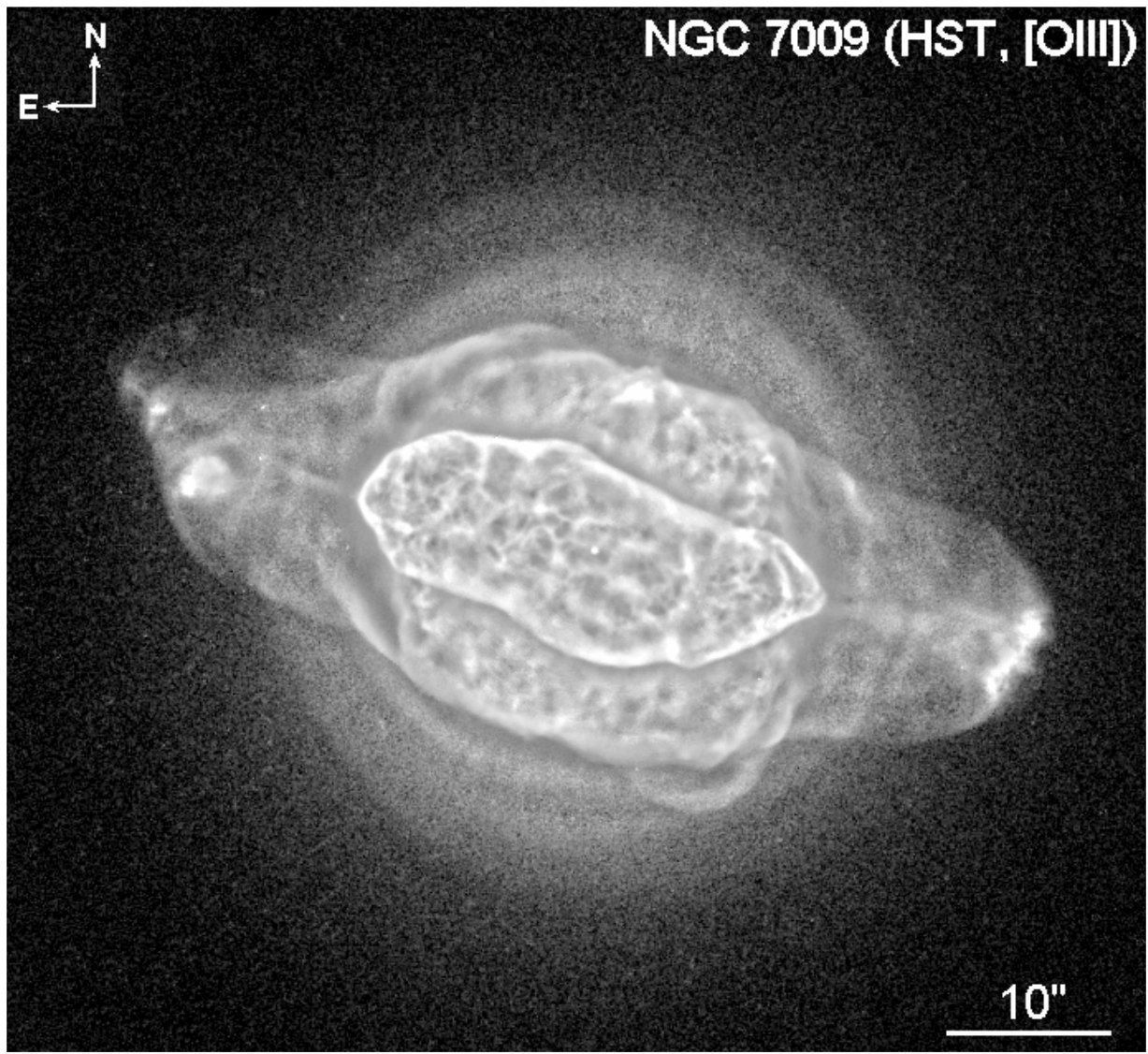

**Figure 17:** An unsharp masked image of NGC 7009 based upon [OIII] observations taken with the HST. Note the bright rim at the limits of the interior elliptical shell, likely indicative of a local shock interface; the bow-shock type structures associated with the external ansae; and the rings within the exterior halo.

A further interesting result relates to the "jet" connecting to the westernmost ansa of the source. This appears (so far as can be discerned) to have a somewhat wedge-shaped appearance, and splays outwards towards the ansa. Such behaviour, where it is confirmed, would be consistent with the modelling of Frank, Balick & Livio (1996), in which they investigate the shock refraction of winds within ellipsoidal nebular shells. It is found that the collision of refracted winds at the major axis limits of the envelopes leads to the formation of jets and ansae similar to those observed here. Where fast wind mass-loss rates decrease with time, and their velocities increase, then this leads to a decrease in jet widths, and increases in their densities and velocities. It may also result in the type of conical structure which is evident in Fig. 17. As a rough gauge to the applicability of such a model, we note that the projected length of the "jet" is $\sim 1.54 \; 10^{12}$ km where the distance is again given by $\sim$ 0.86 kpc. Given a lateral velocity of expansion for the ansae of $\sim 114$ km s$^{-1}$, this then implies an overall lifetime of close to 430 yr. The models of Frank, Balick & Livio (1996) suggest that this is more than enough time to explain the order of magnitude narrowing of the jet in Fig. 17. On the other hand, it is clear that such a model would not be in accord with the hypothesised precession of the easterly ansa. Alternative models for ansa formation have been proposed by Garcia-Segura & Lopez (2000) and Garcia-Segura (1997), in which the structures of PNe depend upon the action of magnetic fields. These analyses would lead to comparable results to those of the previous model, and allow for precession of the ansae and jets. It is interesting to note that the model of Garcia-Segura & Lopez (2000) may also lead to apparent precession effects, even where there is, in reality, no precession within the source.

Finally, there appears to be evidence for at least a couple of rings on either side of the central elliptical shell, with some evidence that the rings also extend into the brighter bow-shock regimes. Although these observations are among the highest resolution results so far obtained for structures of this kind (see also Balick et al. 2001), it seems clear that the rings are featureless and relatively broad, with widths of order 1-3 arcsec. Profiles through the north-westerly and south-westerly portions of these structures are illustrated in Fig. 18, where we have subtracted the underlying components of emission using a fourth order least-squares polynomial fit. It would seem from this that the widths of the rings are by no means uniform, and that there is an appreciable degree of non-circularity, leading to aspect ratios (smallest dimension/largest dimension) of $\sim$ 0.9 or so (see also Fig. 17).

Although the S/N of these results is by no means high, there is also evidence for the kind of sharply defined gradients which might occur through local shock activity. Such shocks have been proposed for NGC 6543, where they may be responsible for enhanced temperatures in the vicinity of the rings (Hyung et al. 2001; although see also Balick et al.

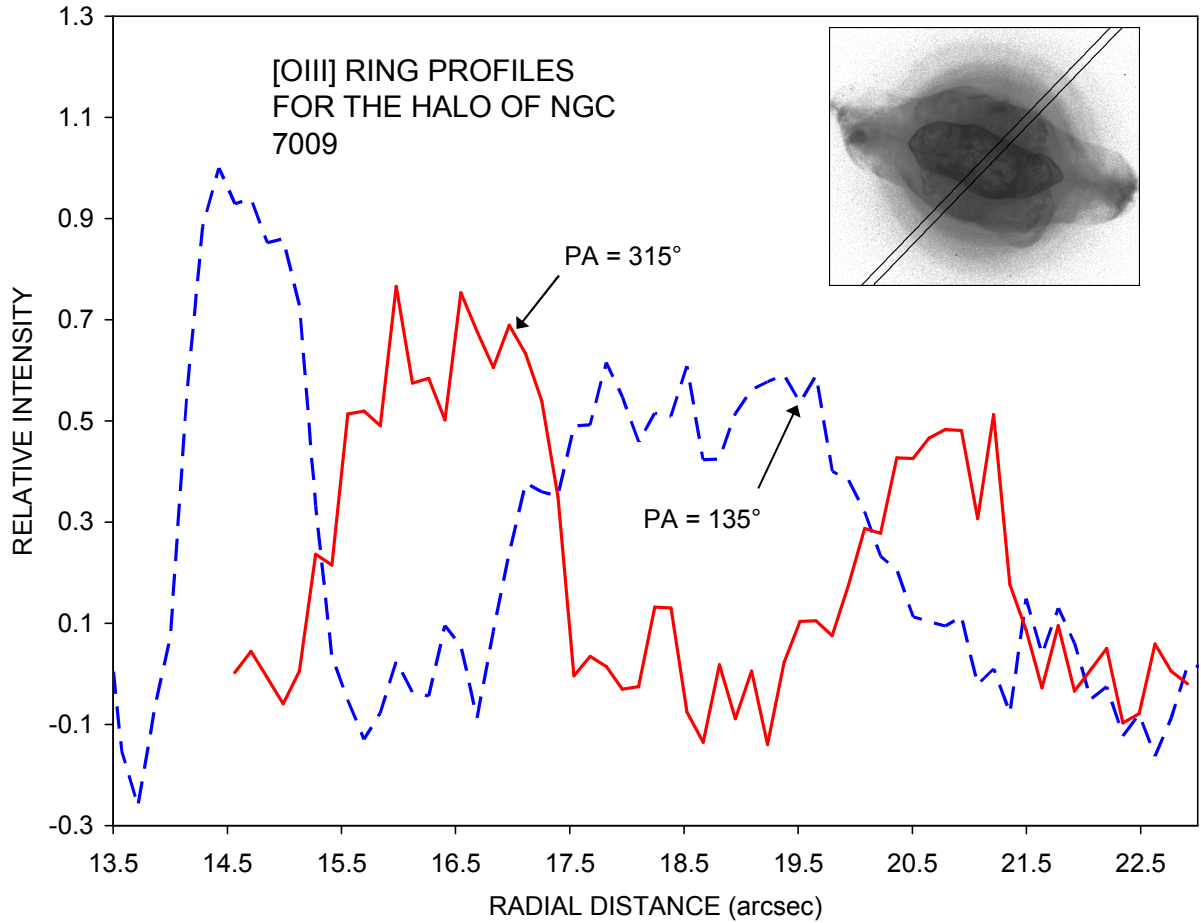

**Figure 18:** Profiles through the north-westerly (PA = 315°) and south-easterly (PA = 135°) halo ring structures, where the width of the slice is 19 pixels ($\equiv$ 1.9 arcsec), and underlying halo emission has been removed using a fourth order least-squares polynomial fit. It would appear that the rings have a slight ellipticity and varying widths.

2001), and in the ring modelling analyses of Mastrodemos & Morris (1999) and Hyung et al. (2001). They may also explain the high excitation nature of this regime noted in Sect. 5.2. By contrast, where this region has been ionised through much of the expansion period of the inner shell (and the absence of any indicators of neutral emission suggests that this may be the case), then thermal expansion at the local velocity of sound (~10 km s$^{-1}$) would be expected to cause greater smearing than is observed. It may of course be that shocks are not continuing to the present day, and that the outer envelope has been flash-ionised in the recent past (see e.g. the analyses of Schönberner & Steffen 2002; Steffen & Schönberner 2003; and Schönberner et al. 2005). Here again, however, the presence of steep gradients in Fig. 18 suggests that shocks may have been involved in their formation.

None of these observations, suggestive through they may be, are entirely decisive in settling this issue, however, and deeper and higher resolution spectroscopy would be of benefit in determining their physical characteristics and outflow kinematics.

## 7. HST Excitation Diagrams, and the Role of Ansae Shocks

It is clear, from the foregoing, that there is evidence for shock excitation of the external ansae, and at the limits of the interior elliptical rim. The ansae appear to form the central portions of more extended bow-shock features, whilst the elliptical rim appears to be shock interacting with the fainter, more circular external shell, leading to narrow decreases and enhancements in high and low excitation emission. The results for the ansae appear in conflict with the ground based spectroscopy, however, and the analysis of line ratios presented in Sect. 5.2. These latter results suggested that none of ratios are consistent with shock enhancement of the emission.

A possible explanation for this discrepancy may be that the stellar radiation field is exceptionally strong, and capable of markedly affecting the ionisation state of pre- and post-shock material (see e.g. Dopita 1997). Similarly, high velocity shocks may, of themselves, appreciably affect the radiation field, and lead to ionisation of the shocks and pre-cursor regimes (Dopita 1995; Shull & McKee 1979) - a situation which occurs at velocities > 100 km s$^{-1}$ similar to those of the present ansae structures. It therefore follows that the radiation characteristics of the post-shock regime may be strongly affected by a variety of factors, and swamped by photo-ionised components of nebular emission.

A final problem in the detection of post-shock cooling radiation is the low spatial resolutions of many ground-based observations. A seeing/slit width of >1-2 arcsec, similar to that of our present spectroscopy, may result in emission from the thin post-shock regimes being completely washed out. The only way of detecting such fluxes would then be through higher resolution observations, and we have therefore used a pixel-by-pixel ratioing of the HST results to undertake a diagnostic analysis of the westerly ansa emission.

The results of this analysis are illustrated in Fig. 19, where we show the variation of Hα/[SII] with Hα/[NII], and of [SII]/Hα with [OIII]/Hα. It is clear, in the case of the upper panel, that the ansa ratios spread into the SNR regime, and this might be taken as evidence for the kind of shock activity surmised from our analysis in Sect. 6. A comparison of these present results with those in Fig. 14, however, reveals that the centroid of the points is significantly different. The mean values <Hα>/<[NII]> and <Hα>/<[SII]> are indicated by the larger yellow diamond, and it would appear that HST results are shifted upwards compared to ground-based spectroscopy. A further correction needs to be applied before these figures can be realistically compared, however. The F656N filter includes both of the nearby [NII] transitions, whilst the wings of the F658N [NII] λ6583 filter include Hα. So, it appears that both of these fluxes, those of [NII] and Hα, are likely to be contaminated by adjacent lines. The system throughputs in the Hα (F656N) filter are of order QT = 0.113 for Hα, 0.034 for [NII] λ6548, and 0.0066 for [NII] λ6584, and similar parameters also apply for the F658N filter (QT = 0.112 for [NII], and 0.0043 for Hα). Given the ratios [NII]/Hα implied by our present results, this would imply that the Hα/[NII] and Hα/[SII] ratios are too high. It follows that the points in Fig. 19 need to be shifted downwards and to the left by ~ 0.09 dex.

A further question relates to the errors in the results – a factor which is extremely variable over the area of the ansa. As an approximate estimate, we determine that Hα fluxes in the weakest portions of the ansa correspond to ~80 electrons/pixel; an estimate that translates into an S/N of ≈ 7.5. Since the Hα exposure has by far the shortest exposure period, and errors in this imaging dominate the uncertainties in line ratios, it follows that the errors in Fig. 19 should be of order ~0.05 dex. The observed scatter in these points is very much larger, however, and this may imply that much of it is real, and reflects intrinsic variations in excitation.

It therefore follows that the difference between the gound- and space-based line ratio estimates appears to be significant, and more extreme than is evident in Figs. 14 & 19; a disparity which may reflect the differences in the regimes which are sampled by these measurements. Whilst the HST results include all of the ansa emission, the ground-based results sample a narrower sliver of this regime. Given that the HST results also extend into the regime of SNRs, then this may indicate that some of the emission is affected by shocks.

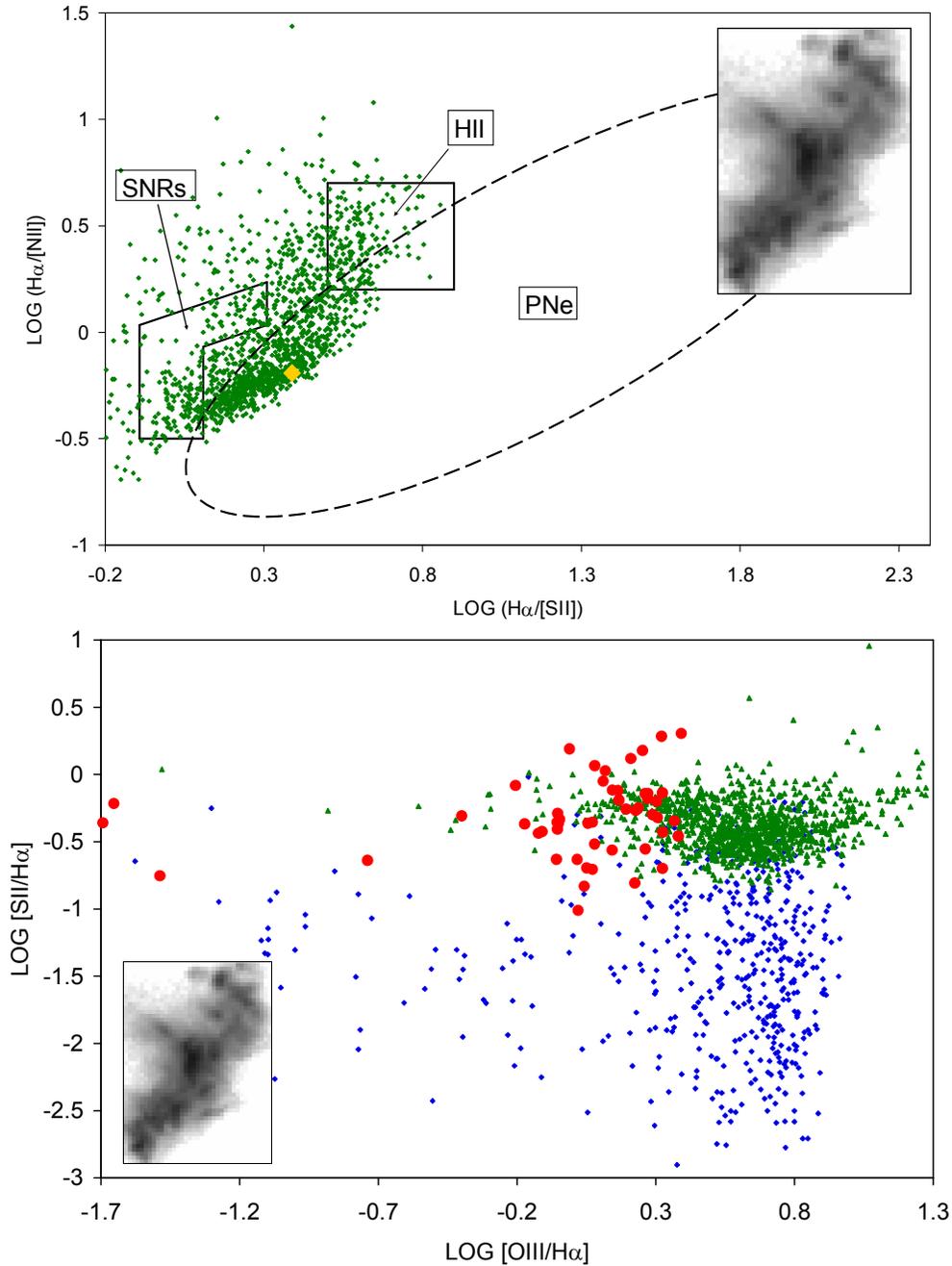

**Figure 19:** The distribution of HST line ratios for the western ansa of NGC 7009, where the area over which the line ratios were obtained is indicated by the inserted figure. The upper panel is the counterpart of the corresponding graph in Fig. 14, in which green symbols correspond to individual pixel line ratios, whilst the lower panel corresponds to a mix of results deriving from the HST ratios (green symbols), the planar- and bow-shock modelling of Hartigan et al.(1987) and Shull & McKee (1979) (large red disks), and ratios for other Galactic PNe (blue symbols) based upon an heterogeneous sample of 492 spectra deriving from Kaler et al. (1997). The latter results have been corrected for reddening using the extinction coefficients of Tylenda et al. (1992), and the extinction curve of Savage & Mathis (1979).

Similar biases afflict the lower panel in Fig. 19, where we illustrate HST results for [OIII]/H$\alpha$ and [SII]/H$\alpha$. We have also included points corresponding to the planar- and bow-shock modelling of Hartigan et al.(1987) and Shull & McKee (1979), and ratios deriving from 492 spectra of Galactic PNe (Kaler et al. 1997). The latter results have been corrected using the extinction coefficients of Tylenda et al. (1992), and the IS extinction curve of Savage & Mathis (1979). It is clear from this that the PNe tend to congregate to the right-hand side of the figure, and the modelling results to the upper centre-left. The two regimes of data points therefore appear to be distinct, and offer a means of distinguishing between photoionisation and shock-excitation processes.

The HST data straddles the two excitation regimes, and suggests that some of the emission is likely to arise from post-shock regimes. Although these results are also affected by contamination of the H$\alpha$ results, any correction for this would shift the points to the upper right-hand side (i.e. to larger values of the ratios). Filter leakage would only account for shifts of ~0.05 dex, however, suggesting that some elements of the emission may derive from shock excitation.

## 8. Conclusions

We have undertaken a multi-wavelength study of the planetary nebula NGC 7009, using visual spectroscopy acquired using the Observatorio Astronomico Nacional (Mexico); making using of MIR spectroscopy and imaging deriving from the ISO and SST; and utilising high resolution narrow band images deriving from the HST. It is clear that the source contains a broad dust continuum longwards of $\lambda \sim 15$ $\mu$m, deriving from grains having a temperature of order ~120 K (for emmisivity exponents $\beta = 0$; the temperature would be lower for larger values of $\beta$). There is also evidence for strong crystalline silicate emission attributable to materials such as Forsterite or ortho- and clino-enstatite. This would be consistent with the C/O ratios > 1 determined from visual ionic line intensities. By contrast, the fluxes observed with the Spitzer IRAC camera appear to be relatively free of dust emission, and are largely dominated by ionic transitions. It is found that surface brightnesses are much stronger at 8.0 μm, where the source appears larger, the morphology of the outer contours is more circular, and the contribution by [ArIII] λ8.991 μm appears to be important. Our mapping also shows that the ratios 5.8μm/4.5μm and 8.0μm/4.5μm are largest in the exterior spherical regime.

The optical spectra are employed to obtain low-resolution mapping of temperatures $T_e$ and densities $n_e$; to determine fluxes for 116 transitions in six sectors of the shell; to illustrate the 2D excitation structure of the source for several transition ratios; and to investigate the possible contributions of shocks in creating the external ansa structures. We confirm that the primary emission from the ansae is consistent with photo-ionisation of local material, although it is possible that evidence for the shocks is being washed out because of low spatial resolutions. A higher resolution study of the HST results concludes that a small fraction of the emission may indeed derive from shocks. Such a conclusion is also in accord with unsharp masked imaging in [OIII], which shows that both of the ansae are associated with bow-shock type structures.

There may also be evidence, in the case of the easterly ansae, for a rotation of the ansa region over time – a variation which may derive from the secular precession of a central collimating engine. The relation between the position angles of the ansa condensations and their distances from the nucleus suggests a rate of angular precession $\sim 5 \times 10^{-2}$ °/yr. It is also possible however that similar effects may arise due to variations in local densities, and/or instabilities in the shocks responsible for these structures.

Finally, we have noted evidence for thin shock-enhanced regions at the limits of the inner elliptical shell, which appear to lead to decreases in [OIII]/Hα, enhancement in [NII]/Hα, and a combination of enhancement and absorption in the case of [SII]/Hα. We also note that the HST [OIII] imaging permits an analysis of the exterior ring system, and that these correspond to the highest resolution observations so far published for structures of this kind. The rings appear to have a typical width of ~1-3 arcsec, are clearly non-circular, and may show evidence for sharp gradients in intensity attributable to local shock compression.

**Acknowledgements**

We would like to thank the referee (Yiannis Tsamis) for several perceptive comments concerning the original script. This work is based, in part, on observations made with the Spitzer Space Telescope, which is operated by the Jet Propulsion Laboratory, California Institute of Technology under a contract with NASA. GRL acknowledges support from CONACyT (Mexico) grant 132671.

# References

Balick B., 2004, AJ, 127, 2262

Balick B., Alexander J., Hajian A.R., Terzian Y., Perinotto M., Patriarchi, P., 1998, ApJ, 116, 360

Balick B., Perinotto M., Maccioni A., Alexander J., Terzian Y., Hajian A.R., 1994, ApJ, 424, 800

Balick B., Rugers M., Terzian Y., Chengalur J.N., 1993, ApJ, 411, 778

Balick B., Wilson J., Hajian A. R., 2001, AJ, 121, 354

Barker T., 1983, ApJ, 267, 630

Barlow M.J., Hales A.S., Storey P.J., Liu X.-W., Tsamis Y.G., Aderin M.E., 2006, in Barlow M.J., Mendez R.H., eds, Proc. IAU Symp. 234, Planetary Nebulae in our Galaxy and Beyond. Cambridge University Press, Cambridge, p. 367

Biretta J., 1996, WFPC2 Instrument Handbook (http://www.stsci.edu/ instruments/wfpc2/ Wfpc2_hand/HTML/W2_1.html)

Bohigas J., Lopez J.A., Aguilar L., 1994, A&A, 291, 595

Bombeck G., Köppen J., Bastian U., 1986, in New Insights in Astrophysics (ESA SP-263), 287

Carey S., Surace J., Glaccum W., Lowrance O., Lacy M., Reach W., 2010, IRAC Instrument Handbook (http://ssc.spitzer.caltech.edu/irac/irac instrument handbook/home/

Cerruti-Sola M., Perinotto M., 1985, ApJ, 291, 247

Cerruti-Sola M., Perinotto M. ,1989, ApJ, 345, 339

Ciardullo R.B., Bond H.E., Sipior M.S., et al., 1999, AJ, 118, 480

Clegg P.E., Ade P.A.R., Armand C. et al. 1996, A&A, 315, L38

Corradi R.L.M., Sánchez-Blázquez P., Mellema G., Gianmanco C., Schwarz H.E., 2004, A&A, 417, 637

Czyzak S.J., Aller L.H., 1979, MNRAS, 188, 229

De Graauw Th., Haser L.N., Beintema D.A. et al. 1996, A&A 315, L49

Dopita M.A., 1977, ApJS, 33, 437

Dopita M.A., 1995, Ap&SS, 233, 215

Dopita M.A., 1997, ApJ, 485, L41

Fernández R., Monteiro H., Schwarz H.E., 2004, ApJ, 603, 595

Frank A., Balick B., Livio M., 1996, ApJ, 471, L53

Garcia-Segura G., 1997, ApJ, 489, L189

Garcia-Segura G., Lopez J.A., 2000, ApJ, 244, 336

Gonçalves D.R., Corradi R.L.M., Mampaso A., Perinotto M., 2003, ApJ, 597, 975

Gonçalves D.R., Ercolano B., Carnero A., Mampaso A., Corradi R.L.M., 2005, MNRAS, 365, 1095

Hartigan P., Raymond J., Hartmann L., 1987, ApJ 316, 323

Huggins P. J., Bachiller R., Cox P., Forveille T., 1996, A&A, 315, 284

Hutsemekers D., Surdej J., 1988, A&A, 219, 237

Hyung S., Aller L.H., 1995a, MNRAS, 273, 958

Hyung S., Aller L.H., 1995b, MNRAS, 273, 973

Hyung S., Fletcher A., Aller L.H., 2003, in Kwok, S., Dopita A., Sutherland R., eds, Proc. IAU Symp. 209, Planetary nebulae; Their Evolution and Role in the Universe. Astron. Soc. Pac., San Francisco, p.401

Hyung S., Mellema G., Lee S.-J., Kim H., 2001, A&A, 378, 587

Jäger C., Molster F.J., Dorschner J., et al., 1998, A&A 339, 904

Kaler J.B., 1976, ApJS, 31, 517

Kaler J.B., Shaw R.S., Browning L.B., 1997, PASP 109, 289

Kastner J.H., Weintraub D.A., Gatley I., Merrill K. M., Probst R.G., 1996, ApJ, 462, 777

Kingsburgh R.L., Barlow M.J., 1992, MNRAS, 257, 317

Koike C., Shibai H., 1998, ISAS report no. 671

Lame N.J., Pogge R.W., 1996, AJ, 111, 2320

Levi L., 1974, Computer Graphics and Image Processing, 3, 163

Liu X.-W., Storey P.J., Barlow M.J., Clegg R.E.S., 1995, MNRAS, 272, 369

Liu X.-W., Storey P.J., Barlow M.J., Danziger I.J., Cohen M., Bryce M., 2000, MNRAS, 312, 585

Luo S.-G., Liu X.-W., Barlow M.J., 2001, MNRAS, 326, 1049

Marigo P., 2008, Mem. S. A. It., 79, 403

Marigo P., Bressan A., Chiosi C., 1998, A&A, 331, 564

Massey P., Strobel K., Barnes J.V., Anderson E., 1988, ApJ, 328, 315

Mastrodemos N., Morris M., 1999, ApJ, 523, 357

Mathis J.S., 1995, Rev. Mex. A&A, 3, 207

Medina J.J, Guerrero M.A., Luridiana V., Miranda L.F., Riera A., Velásquez P.F., 2009, in Corradi R.L.M., Manchado A., Soker N., eds., Asymmetric Planetary Nebulae IV, IAC, Spain, p. 235

Méndez R.H., Kudritzki R.P., Herrero A., 1992, A&A, 260, 329

Molster F., Kemper C., 2005, Space Science Reviews, 119, 3

Ohtani H., 1980, PASJ, 32, 11Fazio G., et al., 2004, ApJS, 154, 10

Perinotto M., Benvenuti P., 1981, A&A, 101, 88

Phillips J. P., Ramos-Larios G., Schröder K.-P., Contreras-Verbena J. L., 2009, MNRAS, 399, 1126

Pittard J.M., Dobson M.S., Durisen R.H., Dyson J.E., Hartquist T.W., O'Brien J.T., 2005, A&A, 438, 11

Polehampton E., 2002, PhD thesis, Univ. Oxford

Reay N.K., Atherton P.D., 1985, MNRAS, 215, 233

Renzini A., Voli M., 1981, A&A, 94, 175

Riesgo H., Lopez J.A., 2006, Rev. Mex. A&A, 42, 47

Rodríguez L.F., Gómez Y., 2007, Rev. Mex. A&A, 43, 173

Rubin R.H., Bhatt N.J., Dufour R.J., et al., 2002, MNRAS, 334, 777

Rubin R.H., Colgan S.W.J., Haas M.R., Lord S.D., Simpson J.P., 1997, ApJ, 479, 332

Sabbadin F., Minello S., Bianchini A. 1977, A&A, 60, 147

Sabbadin F., Turatto M., Capellaro E., Benetti S., Ragazzoni R., 2004, A&A, 416, 955

Savage B.D., Mathis J.S., 1979, ARA&A 17, 73

Schönberner D., Jacob, R., Steffen, M., Perinotto, M., Corradi, R.L., Acker, A., 2005, A&A, 431, 963

Schönberner D., Steffen M., 2002, RevMexAA (Serie de Conferencias), 12, 144

Shaw R.A., Dufour R.J., 1995, PASP, 107, 896

Shortridge K., 1993, in Hanisch R.J., Brissenden R.J.V.,, Narnes J., eds, ASP Conf. Ser. Vol. 52, Astronomical Data Analysis Software and Systems II. Astron. Soc. Pac., San Francisco, p. 219

Shull J.M., McKee C.F., 1979, ApJ 227, 131

Sonneborn G., Iping R.C., Herald J., 2009, in Van Steenberg M.E., Sonneborn G., Moos H.W., Blair W.P., eds, Proc. AIP Conf. Vol. 1135, Future Directions in Ultraviolet Spectroscopy. AIP, Maryland, p. 177

Sonneborn G., Iping R.C., Massa D., Herald J., Chu Y.-H., 2008, in de Koter, A., Smith L.J., Waters L.B.F.M., eds, ASP Conf. Series Vol. 388, Mass Loss from Stars and the Evolution of Stellar Clusters. Astron. Soc. Pac., San Francisco, p. 223

Steffen M., Schönberner D., 2003, in Kwok S., Dopita M., Sutherland R., eds, Proc. IAU Symp. 209, Planetary Nebulae: Their Evolution and Role in the Universe. Astron. Soc. Pac., San Francisco, p. 439

Tinkler C.M., Lamers H.J.G.L.M., 2002, A&A, 384, 987

Tsamis Y.G., Barlow M.J., Liu X.-W., Storey P.J., Danziger I. J., 2004, MNRAS, 353, 953

Tsamis Y.G., Walsh J. R., Péquignot D., Barlow M.J., Danziger I.J., Liu X.-W., 2008, MNRAS, 386, 22

Tylenda R., Acker A., Stenholm B., Köppen, J., 1992, A&AS, 95, 337

van den Hoek L.B., Groenewegen M.A.T., 1997, A&AS 123, 305

TABLE 1

LINE INTENSITIES FOR SIX SEGMENTS OF THE ENVELOPE IN NGC 7009

| ID | λ | T | | I | | II | | III | | IV | | V | | VI | |
|---|---|---|---|---|---|---|---|---|---|---|---|---|---|---|---|
| ION | Å | F/F(Hβ) | σ | F/F(Hβ) | σ | F/F(Hβ) | σ | F/F(Hβ) | σ | F/F(Hβ) | σ | F/F(Hβ) | σ | F/F(Hβ) | σ |
| HI | 4340.465 | 4.86E-01 | 2.84E-02 | 5.03E-01 | 3.60E-02 | 4.88E-01 | 2.59E-02 | 4.44E-01 | 1.95E-02 | 5.02E-01 | 1.05E-01 | 4.55E-01 | 6.12E-02 | 4.34E-01 | 8.13E-03 |
| [OIII] | 4363.191 | 7.94E-02 | 4.89E-04 | 8.54E-02 | 5.34E-04 | 7.91E-02 | 1.01E-03 | 6.47E-02 | 1.24E-03 | 8.90E-02 | 1.52E-02 | 1.02E-01 | 1.01E-02 | 1.90E-01 | 7.32E-03 |
| NII(CII) | 4377.745 | 3.30E-03 | 6.97E-06 | 3.42E-03 | 2.70E-05 | 3.77E-03 | 3.04E-05 | 2.28E-03 | 2.18E-05 | --- | --- | --- | --- | --- | --- |
| HeI | 4387.930 | 7.57E-03 | 2.48E-04 | 7.58E-03 | 2.02E-04 | 7.63E-03 | 2.61E-04 | 7.34E-03 | 2.32E-04 | 1.09E-02 | 1.64E-03 | 8.87E-03 | 2.37E-03 | --- | --- |
| [FeII](OII) | 4413.942 | 2.49E-03 | 1.72E-06 | 3.74E-03 | 3.03E-05 | 2.25E-03 | 2.04E-05 | 1.71E-03 | 3.43E-05 | --- | --- | --- | --- | --- | --- |
| NII | 4433.769 | 2.95E-03 | 1.38E-05 | 8.56E-03 | 2.35E-04 | 2.19E-03 | 1.58E-05 | 1.80E-03 | 3.97E-05 | --- | --- | --- | --- | --- | --- |
| HeI | 4471.499 | 5.19E-02 | 2.97E-03 | 5.05E-02 | 3.49E-03 | 5.22E-02 | 2.73E-03 | 5.08E-02 | 2.09E-03 | 6.07E-02 | 1.18E-02 | 6.02E-02 | 6.41E-03 | 7.54E-02 | 1.33E-02 |
| OII | 4489.468 | 1.46E-03 | 7.06E-05 | 2.99E-03 | 7.40E-04 | 9.60E-04 | 5.45E-04 | 5.30E-04 | 5.20E-04 | --- | --- | --- | --- | --- | --- |
| NIII | 4514.890 | 2.59E-03 | 2.47E-04 | 8.58E-03 | 8.68E-04 | 2.89E-03 | 8.23E-04 | 1.88E-03 | 5.93E-04 | --- | --- | --- | --- | --- | --- |
| NII | 4530.404 | 6.13E-04 | 6.93E-04 | --- | --- | --- | --- | --- | --- | --- | --- | --- | --- | 6.88E-02 | 2.64E-02 |
| HeII | 4541.590 | 7.36E-03 | 2.56E-04 | 9.40E-03 | 8.65E-05 | 6.92E-03 | 3.25E-04 | --- | --- | --- | --- | --- | --- | 1.65E-01 | 5.31E-02 |
| MgI | 4571.161 | 9.60E-04 | 9.50E-05 | 5.55E-04 | 1.59E-04 | 1.43E-03 | 1.02E-04 | 7.20E-04 | 1.08E-04 | --- | --- | --- | --- | --- | --- |
| OII | 4596.171 | 2.48E-03 | 1.30E-04 | 2.58E-03 | 2.62E-04 | 2.91E-03 | 1.66E-04 | 1.81E-03 | 1.88E-04 | --- | --- | --- | --- | --- | --- |
| OII | 4609.436 | 2.05E-03 | 1.13E-05 | 4.79E-03 | 4.65E-04 | 2.31E-03 | 2.89E-05 | 1.60E-03 | 5.54E-05 | --- | --- | --- | --- | --- | --- |
| CI(SII, NII) | 4621.570 | --- | --- | --- | --- | --- | --- | --- | --- | --- | --- | 3.27E-03 | 5.17E-04 | --- | --- |
| NIII | 4634.079 | 5.49E-03 | 1.79E-04 | 5.99E-03 | 1.81E-04 | 5.94E-03 | 1.71E-04 | 4.34E-03 | 3.13E-05 | 1.28E-02 | 2.30E-03 | 2.15E-02 | 1.84E-03 | --- | --- |
| NIII | 4640.604 | 5.04E-02 | 3.18E-03 | 6.95E-02 | 5.04E-03 | 4.86E-02 | 2.84E-03 | 2.79E-02 | 1.25E-03 | 9.11E-03 | 2.44E-03 | 7.83E-03 | 1.25E-03 | --- | --- |
| OII | 4650.825 | 9.60E-03 | 4.20E-04 | 1.01E-02 | 3.44E-04 | 9.98E-03 | 4.02E-04 | 7.71E-03 | 2.30E-04 | 8.91E-03 | 3.81E-04 | 1.37E-02 | 4.74E-03 | --- | --- |
| OII | 4661.625 | 3.25E-03 | 9.04E-06 | 4.74E-03 | 2.01E-04 | 3.11E-03 | 7.47E-06 | 2.43E-03 | 5.72E-06 | 1.93E-03 | 2.11E-04 | 1.30E-02 | 9.57E-03 | --- | --- |
| HeII | 4685.680 | 1.92E-01 | 1.23E-02 | 3.03E-01 | 2.33E-02 | 1.89E-01 | 1.10E-02 | 4.02E-02 | 1.77E-03 | 2.04E-02 | 3.53E-03 | 2.64E-02 | 3.76E-03 | 6.52E-02 | 9.14E-03 |
| [ArIV] | 4711.352 | 5.16E-02 | 2.30E-03 | 6.35E-02 | 2.55E-03 | 4.92E-02 | 1.98E-03 | 4.25E-02 | 1.95E-03 | 5.07E-02 | 1.02E-02 | 3.52E-02 | 2.54E-03 | 8.07E-02 | 5.41E-03 |
| [ArIV] | 4740.205 | 4.92E-02 | 2.19E-03 | 6.33E-02 | 2.60E-03 | 4.72E-02 | 1.94E-03 | 3.52E-02 | 1.57E-03 | 3.45E-02 | 6.76E-03 | 2.51E-02 | 1.86E-03 | 9.88E-02 | 8.62E-03 |
| CII | 4802.454 | 2.42E-04 | 1.34E-05 | 2.93E-04 | 5.72E-05 | 2.32E-04 | 9.45E-06 | --- | --- | --- | --- | --- | --- | --- | --- |
| HI | 4861.327 | 1.00E+00 | --- | 1.00E+00 | --- | 1.00E+00 | --- | 1.00E+00 | --- | 1.00E+00 | --- | 1.00E+00 | --- | 1.00E+00 | --- |
| [OII] | 4906.700 | 1.80E-03 | 2.13E-04 | 2.80E-03 | 6.98E-04 | 1.63E-03 | 1.51E-04 | 2.06E-03 | 2.85E-04 | 2.13E-02 | 9.76E-03 | 2.92E-03 | 1.01E-03 | 3.37E-02 | 1.10E-02 |
| HeI | 4921.933 | 1.51E-02 | 9.03E-04 | 1.48E-02 | 9.33E-04 | 1.53E-02 | 8.62E-04 | 1.47E-02 | 6.51E-04 | 1.89E-02 | 2.51E-03 | 2.13E-02 | 1.20E-03 | 4.04E-02 | 9.77E-03 |
| [OIII] | 4931.229 | 1.36E-03 | 1.61E-05 | 1.92E-03 | 4.23E-05 | 1.28E-03 | 1.31E-05 | 1.17E-03 | 3.39E-05 | 5.54E-03 | 5.08E-05 | 4.51E-03 | 7.27E-04 | --- | --- |
| [OIII] | 4958.915 | 3.60E+00 | 1.92E-01 | 3.56E+00 | 2.28E-01 | 3.65E+00 | 1.75E-01 | 3.32E+00 | 1.30E-01 | 4.19E+00 | 6.95E-01 | 4.32E+00 | 4.87E-01 | 5.58E+00 | 2.89E-01 |
| [OIII] | 5006.845 | 1.09E+01 | 7.11E-01 | 1.10E+01 | 8.64E-01 | 1.11E+01 | 6.63E-01 | 9.26E+00 | 4.58E-01 | 1.61E+01 | 3.26E+00 | 1.66E+01 | 2.26E+00 | 1.67E+01 | 9.86E-01 |
| SiII | 5041.022 | 1.09E-03 | 9.29E-07 | 1.18E-03 | 1.23E-05 | 1.27E-03 | 3.12E-07 | 6.56E-04 | 5.43E-05 | --- | --- | --- | --- | --- | --- |
| HeII | 5047.741 | 1.36E-03 | 1.84E-05 | 1.34E-03 | 3.04E-05 | 1.43E-03 | 1.20E-05 | 1.07E-03 | 3.05E-05 | --- | --- | --- | --- | --- | --- |
| SiII | 5056.067 | 5.84E-04 | 8.21E-06 | 8.46E-04 | 3.35E-07 | 6.17E-04 | 1.01E-05 | --- | --- | --- | --- | --- | --- | --- | --- |
| CII | 5121.850 | 2.51E-04 | 3.40E-05 | 9.38E-04 | 1.05E-04 | 1.95E-04 | 5.14E-06 | --- | --- | --- | --- | --- | --- | --- | --- |
| OI | 5130.530 | 4.88E-04 | 1.31E-05 | 3.06E-03 | 1.08E-05 | 5.09E-04 | 1.48E-05 | 4.39E-04 | 2.47E-05 | --- | --- | --- | --- | --- | --- |
| OI | 5146.692 | 6.26E-04 | 1.91E-06 | 1.53E-03 | 2.47E-05 | 6.64E-04 | 2.40E-05 | 4.53E-04 | 1.96E-05 | --- | --- | --- | --- | --- | --- |

TABLE 1 (cont.)

| ID | λ | T | | I | | II | | III | | IV | | V | | VI | |
|---|---|---|---|---|---|---|---|---|---|---|---|---|---|---|---|
| ION | Å | F/F(Hβ) | σ | F/F(Hβ) | σ | F/F(Hβ) | σ | F/F(Hβ) | σ | F/F(Hβ) | σ | F/F(Hβ) | σ | F/F(Hβ) | σ |
| [FeII] | 5158.858 | --- | --- | --- | --- | 9.87E-05 | 1.02E-05 | --- | --- | --- | --- | --- | --- | --- | --- |
| NI | 5173.564 | --- | --- | --- | --- | 2.53E-04 | 1.08E-04 | --- | --- | --- | --- | --- | --- | --- | --- |
| NI | 5179.521 | --- | --- | 2.26E-04 | 5.28E-05 | 2.88E-04 | 1.12E-04 | --- | --- | --- | --- | --- | --- | --- | --- |
| [ArIII] | 5191.702 | 9.07E-04 | 3.72E-06 | 7.31E-04 | 1.43E-05 | 8.83E-04 | 3.10E-05 | 9.37E-04 | 1.47E-05 | --- | --- | --- | --- | --- | --- |
| [NI] | 5198.012 | 3.38E-04 | 2.95E-05 | --- | --- | 5.24E-04 | 8.38E-06 | --- | --- | --- | --- | --- | --- | --- | --- |
| [FeIII] | 5270.562 | 6.30E-04 | 9.34E-06 | 3.48E-04 | 2.46E-05 | 5.27E-04 | 1.66E-05 | 3.88E-04 | 6.61E-06 | --- | --- | 3.30E-02 | 3.55E-03 | --- | --- |
| --- | 5293.000 | 9.02E-04 | 2.03E-05 | 2.81E-03 | 6.33E-05 | --- | --- | --- | --- | --- | --- | 3.27E-03 | 4.12E-04 | --- | --- |
| OIV | 5305.300 | 4.36E-04 | 8.25E-06 | 1.80E-03 | 1.18E-05 | 1.38E-04 | 8.12E-06 | --- | --- | --- | --- | 4.60E-03 | 6.61E-04 | --- | --- |
| [ClIV] | 5323.270 | 2.99E-04 | 2.08E-05 | 8.60E-04 | 4.95E-05 | 1.64E-04 | 4.89E-06 | --- | --- | --- | --- | --- | --- | --- | --- |
| CII | 5332.771 | --- | --- | 5.00E-04 | 6.46E-05 | --- | --- | --- | --- | --- | --- | --- | --- | --- | --- |
| CII | 5342.392 | 9.29E-04 | 1.85E-05 | 1.45E-03 | 1.41E-05 | 7.44E-04 | 2.51E-05 | 5.48E-04 | 7.78E-06 | --- | --- | --- | --- | --- | --- |
| FeI | 5358.100 | 2.22E-04 | 3.10E-05 | 7.30E-04 | 9.14E-05 | 1.40E-04 | 1.76E-05 | 1.01E-04 | 2.12E-05 | --- | --- | --- | --- | --- | --- |
| CII | 5368.205 | 3.83E-04 | 5.53E-05 | --- | --- | 4.17E-04 | 2.73E-05 | 2.00E-04 | 1.16E-04 | --- | --- | --- | --- | --- | --- |
| [FeII] | 5376.691 | 4.03E-04 | 5.77E-05 | 8.60E-04 | 2.58E-04 | 2.95E-04 | 3.53E-05 | 5.55E-04 | 1.81E-04 | --- | --- | --- | --- | --- | --- |
| HeII | 5411.520 | 1.57E-02 | 9.91E-04 | 2.32E-02 | 1.73E-03 | 1.56E-02 | 9.16E-04 | 4.94E-03 | 2.03E-04 | --- | --- | --- | --- | --- | --- |
| SII | 5454.035 | 1.94E-04 | 1.19E-04 | 1.34E-04 | 2.36E-04 | 1.59E-04 | 9.25E-06 | 8.77E-05 | 4.79E-05 | --- | --- | --- | --- | --- | --- |
| NII | 5480.050 | 9.53E-05 | 9.89E-06 | --- | --- | 4.53E-05 | 1.33E-05 | --- | --- | --- | --- | --- | --- | --- | --- |
| OI | 5486.600 | 2.10E-04 | 2.13E-06 | 3.96E-04 | 2.61E-05 | 1.33E-04 | 1.06E-05 | 1.65E-05 | --- | --- | --- | --- | --- | --- | --- |
| NII([FeII]) | 5495.655 | 1.07E-04 | 1.44E-05 | --- | --- | --- | --- | --- | --- | --- | --- | --- | --- | --- | --- |
| [ClIII] | 5517.686 | 4.89E-03 | 2.70E-04 | 4.15E-03 | 2.83E-04 | 5.00E-03 | 2.58E-04 | 2.08E-04 | 7.95E-03 | 1.20E-03 | 9.71E-03 | 6.28E-04 | --- | --- | | |
| [ClIII] | 5537.853 | 6.24E-03 | 3.66E-04 | 5.79E-03 | 4.16E-04 | 6.46E-03 | 3.55E-04 | 2.44E-04 | 6.72E-03 | 9.68E-04 | 8.48E-03 | 4.85E-04 | --- | --- | | |
| NII | 5551.930 | 2.71E-04 | 1.26E-04 | 2.64E-04 | 1.54E-04 | 1.44E-04 | 7.00E-05 | 1.27E-05 | --- | --- | --- | --- | --- | --- | | |
| NII | 5565.300 | 8.42E-05 | 3.51E-05 | 1.53E-04 | 1.00E-04 | 5.65E-05 | 5.03E-05 | 2.37E-05 | --- | --- | 4.51E-03 | 3.73E-05 | --- | --- | | |
| [NII] | 5666.630 | 1.16E-03 | 8.75E-05 | 1.95E-03 | 5.34E-05 | 9.38E-04 | 2.09E-05 | 2.74E-05 | --- | --- | --- | --- | --- | --- | | |
| NII ([FeVI]) | 5679.552 | 2.05E-03 | 2.05E-05 | 2.13E-03 | 1.04E-04 | 2.03E-03 | 8.25E-05 | 3.35E-06 | 2.26E-03 | 3.48E-04 | --- | --- | --- | --- | | |
| SiII | 5689.180 | 6.53E-04 | 8.24E-05 | 3.84E-04 | 6.02E-05 | 3.97E-04 | 2.80E-05 | 3.18E-05 | --- | --- | --- | --- | --- | --- | | |
| FeI | 5701.550 | 4.03E-04 | 3.69E-05 | 2.30E-04 | 4.02E-05 | 3.16E-04 | 1.27E-05 | 3.64E-05 | --- | --- | --- | --- | --- | --- | | |
| NII | 5710.763 | 2.77E-04 | 2.30E-05 | 2.15E-04 | 3.09E-05 | 1.64E-04 | 1.55E-05 | 2.37E-05 | --- | --- | --- | --- | --- | --- | | |
| [FeVI] | 5721.100 | 1.21E-04 | 4.86E-05 | 7.35E-04 | 9.42E-05 | 1.19E-04 | 3.86E-05 | 2.74E-05 | --- | --- | --- | --- | --- | --- | | |
| [NII] | 5754.629 | 2.89E-03 | 1.64E-04 | 2.29E-03 | 1.02E-04 | 3.31E-03 | 1.87E-04 | 2.67E-05 | 3.65E-03 | 3.24E-04 | 3.55E-02 | 4.14E-03 | --- | --- | | |
| NII | 5801.510 | 1.00E-03 | 4.33E-05 | 2.65E-03 | 1.83E-04 | 5.90E-04 | 1.62E-05 | 8.35E-05 | --- | --- | --- | --- | --- | --- | | |
| CIV | 5812.140 | 7.39E-04 | 2.87E-05 | 1.93E-03 | 1.19E-04 | 5.30E-04 | 1.39E-05 | 1.19E-06 | --- | --- | --- | --- | --- | --- | | |
| CIV | 5820.610 | 2.11E-04 | 7.03E-06 | 3.11E-04 | 1.54E-05 | 1.96E-04 | 7.47E-06 | 3.52E-06 | --- | --- | --- | --- | --- | --- | | |
| HeII | 5837.060 | 4.91E-05 | 1.40E-05 | 2.78E-04 | 2.18E-05 | --- | --- | 2.74E-05 | --- | --- | --- | --- | --- | --- | | |
| HeII | 5846.704 | 5.60E-05 | 1.65E-05 | 1.87E-04 | 2.61E-05 | 9.78E-05 | 9.08E-06 | 1.88E-05 | --- | --- | --- | --- | --- | --- | | |
| --- | 5859.582 | 2.02E-04 | 5.15E-05 | --- | --- | 2.09E-04 | 5.48E-05 | --- | --- | --- | --- | --- | --- | --- | --- |

TABLE 1 (cont.)

| ID | λ | T | | I | | II | | III | | IV | | V | | VI | |
|---|---|---|---|---|---|---|---|---|---|---|---|---|---|---|---|
| ION | Å | F/F(Hβ) | σ | F/F(Hβ) | σ | F/F(Hβ) | σ | F/F(Hβ) | σ | F/F(Hβ) | σ | F/F(Hβ) | σ | F/F(Hβ) | σ |
| HeI | 5875.650 | 1.58E-01 | 1.05E-02 | 1.52E-01 | 1.22E-02 | 1.58E-01 | 9.71E-03 | 1.65E-01 | 8.29E-03 | 1.59E-01 | 3.31E-02 | 1.55E-01 | 2.04E-02 | 1.60E-01 | 2.88E-03 |
| FeI | 5904.670 | 3.79E-04 | 4.20E-05 | 9.95E-04 | 4.75E-05 | 2.38E-04 | 4.86E-05 | 1.94E-04 | 4.95E-05 | --- | --- | --- | --- | --- | --- |
| HeII | 5912.900 | 2.17E-04 | 3.70E-05 | 8.06E-04 | 2.85E-05 | 1.51E-04 | 3.22E-05 | 6.06E-05 | 3.55E-05 | --- | --- | --- | --- | --- | --- |
| NII | 5931.773 | 3.49E-04 | 1.99E-05 | 8.41E-04 | 1.30E-06 | 3.06E-04 | 2.01E-05 | 7.62E-05 | 3.50E-05 | --- | --- | --- | --- | --- | --- |
| NII | 5941.663 | 1.78E-04 | 2.97E-05 | 4.31E-04 | 2.96E-05 | 2.15E-04 | 2.34E-05 | --- | --- | --- | --- | --- | --- | --- | --- |
| HeII(NII) | 5952.850 | 3.19E-04 | 2.11E-05 | 6.22E-04 | 1.11E-05 | 2.95E-04 | 2.11E-05 | 1.05E-04 | 3.91E-05 | --- | --- | --- | --- | --- | --- |
| HeII | 5977.700 | 5.57E-04 | 2.96E-05 | 1.26E-03 | 2.01E-05 | 5.39E-04 | 5.99E-05 | --- | --- | --- | --- | --- | --- | --- | --- |
| HeII | 6004.800 | 3.66E-04 | 2.56E-05 | 5.82E-04 | 4.37E-05 | 3.27E-04 | 7.55E-06 | --- | --- | --- | --- | --- | --- | --- | --- |
| CI | 6019.822 | --- | --- | --- | --- | --- | --- | --- | --- | --- | --- | --- | --- | --- | --- |
| HeII | 6037.200 | 1.82E-04 | 1.93E-05 | 3.13E-04 | 4.16E-05 | 2.36E-04 | 5.70E-06 | --- | --- | --- | --- | --- | --- | --- | --- |
| NeI | 6074.324 | 4.86E-04 | 2.16E-05 | 7.71E-04 | 5.09E-05 | 4.50E-04 | 5.75E-06 | --- | --- | --- | --- | --- | --- | --- | --- |
| [FeI] | 6082.720 | 2.39E-04 | 2.33E-05 | 3.00E-04 | 5.39E-05 | 2.16E-04 | 1.29E-05 | --- | --- | --- | --- | --- | --- | --- | --- |
| HeII | 6102.000 | 1.97E-03 | 6.76E-05 | 2.88E-03 | 1.82E-04 | 1.94E-03 | 8.48E-05 | 2.05E-03 | 4.19E-05 | --- | --- | --- | --- | --- | --- |
| [KIV] | 6118.300 | 5.80E-04 | 3.63E-05 | 1.59E-03 | 4.86E-05 | 6.84E-04 | 1.84E-06 | --- | --- | --- | --- | --- | --- | --- | --- |
| HeII | 6153.000 | 7.64E-04 | 3.52E-05 | 1.67E-03 | 5.25E-05 | 9.43E-04 | 6.32E-06 | --- | --- | --- | --- | --- | --- | --- | --- |
| --- | 6170.170 | 5.87E-04 | 3.20E-05 | 7.77E-04 | 4.33E-07 | 6.87E-04 | 8.54E-06 | --- | --- | --- | --- | --- | --- | --- | --- |
| NII | 6191.560 | --- | --- | 5.10E-03 | 9.84E-05 | 9.55E-04 | 6.52E-05 | --- | --- | --- | --- | --- | --- | --- | --- |
| [FeI] | 6234.000 | 7.44E-04 | 2.26E-05 | 1.20E-03 | 6.47E-05 | 8.24E-04 | 7.25E-06 | 3.69E-04 | 5.69E-05 | --- | --- | --- | --- | --- | --- |
| HeII | 6233.880 | --- | --- | 2.38E-03 | 9.45E-05 | --- | --- | --- | --- | --- | --- | --- | --- | --- | --- |
| [OI] | 6300.404 | 2.72E-03 | 1.29E-04 | --- | --- | 3.52E-03 | 1.65E-04 | 1.45E-03 | 1.26E-05 | 4.68E-03 | 6.10E-04 | 1.43E-01 | 1.90E-02 | 1.43E+01 | 5.20E+00 |
| [SIII] | 6312.107 | 1.38E-02 | 9.19E-04 | 1.28E-02 | 9.21E-04 | 1.46E-02 | 8.99E-04 | 1.21E-02 | 5.84E-04 | 1.32E-02 | 2.37E-03 | 1.73E-02 | 1.06E-03 | --- | --- |
| [FeII] | 6325.191 | 2.16E-04 | 4.15E-05 | 1.47E-04 | --- | 1.47E-04 | 4.88E-05 | --- | --- | 1.02E-03 | 2.51E-04 | --- | --- | --- | --- |
| SiI | 6347.193 | 5.60E-04 | 2.60E-05 | 4.95E-04 | 9.79E-05 | 7.20E-04 | 2.87E-05 | 4.04E-04 | 5.21E-05 | --- | --- | --- | --- | --- | --- |
| [OI] | 6363.886 | 9.42E-04 | 3.06E-05 | --- | --- | 1.25E-03 | 3.34E-05 | 5.39E-04 | 7.15E-05 | --- | --- | 4.38E-02 | 4.60E-03 | --- | --- |
| SiI | 6371.418 | 6.85E-04 | 3.94E-05 | 9.15E-04 | 1.33E-04 | 7.71E-04 | 6.43E-05 | 4.96E-04 | 7.15E-05 | --- | --- | --- | --- | --- | --- |
| [MnV](ClI) | 6393.660 | 4.61E-04 | 9.78E-06 | 3.64E-04 | 4.30E-05 | 4.12E-04 | 2.38E-06 | 4.61E-04 | 2.53E-05 | --- | --- | --- | --- | --- | --- |
| HeII | 6406.380 | 8.93E-04 | 2.23E-05 | 1.18E-03 | 7.06E-05 | 8.45E-04 | 3.00E-05 | 3.11E-04 | 2.55E-05 | --- | --- | --- | --- | --- | --- |
| [ArV] | 6434.720 | 7.32E-04 | 3.50E-05 | 9.73E-04 | 1.66E-05 | 5.01E-04 | 5.52E-05 | 4.70E-04 | 8.85E-05 | --- | --- | --- | --- | --- | --- |
| CII | 6461.848 | 1.09E-03 | 7.19E-05 | 1.38E-03 | 5.71E-05 | 9.01E-04 | 6.70E-06 | 7.67E-04 | 5.24E-05 | --- | --- | --- | --- | --- | --- |
| --- | 6483.000 | --- | --- | 6.48E-04 | 9.40E-05 | 1.54E-03 | 4.89E-04 | --- | --- | --- | --- | --- | --- | --- | --- |
| NI([NII]) | 6484.880 | 2.17E-03 | 6.14E-05 | 3.09E-03 | 6.95E-05 | 4.02E-04 | 1.04E-04 | 1.05E-03 | 1.13E-04 | --- | --- | --- | --- | --- | --- |
| ArII | 6500.280 | 7.90E-04 | 7.54E-05 | 1.93E-03 | 7.51E-05 | 7.18E-04 | 8.92E-05 | --- | --- | --- | --- | --- | --- | --- | --- |
| FeI | 6509.560 | 1.04E-03 | 1.47E-04 | 5.14E-04 | 2.53E-06 | 6.66E-04 | 8.59E-05 | 6.96E-04 | 2.80E-04 | --- | --- | --- | --- | --- | --- |
| NII(HeII) | 6527.257 | 1.32E-03 | 1.19E-04 | 1.92E-03 | 1.07E-05 | 1.31E-03 | 4.68E-05 | --- | --- | --- | --- | --- | --- | --- | --- |
| [NII] | 6548.096 | 5.85E-02 | 3.01E-02 | 1.98E-02 | 8.51E-04 | 2.56E-02 | 2.82E-03 | 2.94E-02 | 9.05E-03 | 7.05E-02 | 7.20E-03 | 6.37E-01 | 8.40E-02 | --- | --- |

TABLE 1 (cont.)

| ID | λ | T | | I | | II | | III | | IV | | V | | VI | |
|---|---|---|---|---|---|---|---|---|---|---|---|---|---|---|---|
| ION | Å | F/F(Hβ) | σ | F/F(Hβ) | σ | F/F(Hβ) | σ | F/F(Hβ) | σ | F/F(Hβ) | σ | F/F(Hβ) | σ | F/F(Hβ) | σ |
| HI | 6562.804 | 2.79E+00 | 1.71E-01 | 2.82E+00 | 2.21E-01 | 2.73E+00 | 1.52E-01 | 2.96E+00 | 1.50E-01 | 2.99E+00 | 6.39E-01 | 2.77E+00 | 3.92E-01 | 2.92E+00 | 1.63E-01 |
| [NII] | 6583.467 | 1.26E-01 | 6.63E-03 | 5.43E-02 | 1.49E-02 | 1.51E-01 | 7.11E-03 | 9.94E-02 | 6.57E-03 | 2.19E-01 | 3.96E-02 | 2.16E+00 | 3.04E-01 | 1.17E-01 | 1.27E-02 |
| --- | 6619.000 | 5.01E-04 | 2.78E-04 | 5.41E-04 | 3.99E-05 | 3.31E-04 | 5.12E-05 | --- | --- | --- | --- | --- | --- | --- | --- |
| NI | 6644.960 | 4.65E-04 | 1.29E-04 | 1.08E-03 | 6.51E-06 | 3.22E-04 | 1.57E-05 | --- | --- | --- | --- | --- | --- | --- | --- |
| HeI | 6678.153 | 4.53E-02 | 3.06E-03 | 4.50E-02 | 3.61E-03 | 4.51E-02 | 2.82E-03 | 4.58E-02 | 2.33E-03 | 4.51E-02 | 8.96E-03 | 3.77E-02 | 3.35E-03 | 3.48E-02 | 9.13E-03 |
| [SII] | 6716.523 | 1.22E-02 | 7.14E-04 | 5.74E-03 | 2.36E-04 | 1.42E-02 | 7.78E-04 | 1.05E-02 | 3.82E-04 | 3.26E-02 | 6.21E-03 | 2.32E-01 | 3.13E-02 | --- | --- |
| [SII] | 6730.893 | 1.84E-02 | 1.16E-03 | 6.72E-03 | 3.50E-04 | 2.23E-02 | 1.31E-03 | 1.51E-02 | 6.32E-04 | 3.73E-02 | 7.24E-03 | 2.84E-01 | 3.86E-02 | --- | --- |
| HeI | 6769.672 | 2.33E-03 | 6.33E-05 | 4.70E-03 | 8.56E-05 | 2.31E-03 | 1.28E-04 | 1.84E-03 | 1.71E-04 | --- | --- | --- | --- | --- | --- |
| [KIV] | 6797.000 | 7.00E-04 | 1.11E-05 | 6.99E-04 | 1.61E-05 | 7.35E-04 | 5.09E-06 | 7.15E-04 | 6.29E-05 | --- | --- | --- | --- | --- | --- |
| NII | 6809.953 | 3.99E-04 | 5.72E-05 | --- | --- | 3.87E-04 | 1.04E-04 | --- | --- | --- | --- | --- | --- | --- | --- |
| --- | 6826.000 | 1.37E-03 | 6.14E-05 | 1.92E-03 | 5.08E-04 | 4.86E-04 | 3.66E-04 | 1.24E-03 | 1.14E-04 | --- | --- | --- | --- | --- | --- |
| F(Hβ) | $10^{-16}$ erg/cm$^2$/s | 4.75E+05 | 3.38E+04 | 7.42E+04 | 6.31E+03 | 3.00E+05 | 1.98E+04 | 7.73E+04 | 4.23E+03 | 5.45E+03 | 1.18E+03 | 3.68E+03 | 5.30E+02 | 9.18E+02 | 5.74E+01 |